%% file: SUS-16-038_temp.tex
\pdfoutput=1

\documentclass[11pt,twoside,a4paper,cmspaper,final,collab]{cms-tdr}
\def\svnVersion{461107P}\def\cmsCernNoTag{CERN-EP-2018-003}\def\cmsCernDate{\today}\def\cmsMessage{Published in the Journal of High Energy Physics as \href{http://dx.doi.org/10.1007/JHEP05(2018)025}{\doi{10.1007/JHEP05(2018)025.}}}

\begin{document}\cmsNoteHeader{SUS-16-038}

\hyphenation{had-ron-i-za-tion}
\hyphenation{cal-or-i-me-ter}
\hyphenation{de-vices}
\RCS$HeadURL: svn+ssh://svn.cern.ch/reps/tdr2/papers/SUS-16-038/trunk/SUS-16-038.tex $
\RCS$Id: SUS-16-038.tex 461079 2018-05-21 16:39:43Z bainbrid $

\providecommand{\CL}{CL\xspace}
\providecommand{\NA}{\ensuremath{\text{---}}}
\newcommand{\ph}[1]{\phantom{#1}}
\newcommand{\tf}{\ensuremath{\mathcal{R}}\xspace}
\newcommand{\cls}{\ensuremath{\text{\CL}_\text{s}}\xspace}
\newcommand{\jet}[1]{\ensuremath{\mathrm{j}_\text{#1}}\xspace}
\newcommand{\fh}{\ensuremath{f_{\mathrm{h}^{\pm}}}\xspace}
\newcommand{\fhleadjet}{\ensuremath{f_{\mathrm{h}^{\pm}}^{\jet{1}}}\xspace}
\newcommand{\ate}{\ensuremath{\mathcal{A}\varepsilon}\xspace}
\newcommand{\ctau}{\ensuremath{c\tau_{0}}\xspace}
\newcommand{\um}{\ensuremath{\,{\mu}\mathrm{m}}\xspace}
\newcommand{\njet}{\ensuremath{n_{\text{jet}}}\xspace}
\newcommand{\nb}{\ensuremath{n_{\mathrm{b}}}\xspace}
\newcommand{\scalht}{\ensuremath{H_{\mathrm{T}}}\xspace}
\newcommand{\alphat}{\ensuremath{\alpha_{\mathrm{T}}}\xspace}
\newcommand{\bdphi}{\ensuremath{\Delta\phi^{*}_\text{min}}\xspace}
\newcommand{\bdphimod}{\ensuremath{\Delta\phi^{*_{\, 25}}_\text{min}}\xspace}
\newcommand{\mhtmet}{\ensuremath{\mht / \ptmiss}\xspace}
\newcommand{\jets}{\ensuremath{\text{jets}}}
\newcommand{\mj}{\ensuremath{\mu\text{+\jets}}\xspace}
\newcommand{\mmj}{\ensuremath{\mu\mu\text{+\jets}}\xspace}
\newcommand{\gj}{\ensuremath{\gamma\text{+\jets}}\xspace}
\newcommand{\znunu}{\ensuremath{\cPZ \to \cPgn\cPagn}\xspace}
\newcommand{\znunuj}{\ensuremath{\cPZ(\to \cPgn\cPagn)\text{+\jets}}\xspace}
\newcommand{\zj}{\ensuremath{\cPZ\text{+\jets}}\xspace}
\newcommand{\wj}{\ensuremath{\PW\text{+\jets}}\xspace}
\newcommand{\wmj}{\ensuremath{\PW(\to\mu\nu)\text{+\jets}}\xspace}
\newcommand{\lost}{\ensuremath{\ell_\text{lost}}\xspace}

\cmsNoteHeader{SUS-16-038}

\title{Search for natural and split supersymmetry in proton-proton
  collisions at $\sqrt{s} = 13\TeV$ in final states with jets and
  missing transverse momentum}

\date{\today}

\abstract{A search for supersymmetry (SUSY) is performed in final
  states comprising one or more jets and missing transverse momentum
  using data from proton-proton collisions at a centre-of-mass energy
  of 13\TeV. The data were recorded with the CMS detector at the CERN
  LHC in 2016 and correspond to an integrated luminosity of
  35.9\fbinv. The number of signal events is found to agree with the
  expected background yields from standard model processes. The
  results are interpreted in the context of simplified models of SUSY
  that assume the production of gluino or squark pairs and their
  prompt decay to quarks and the lightest neutralino.  The masses of
  bottom, top, and mass-degenerate light-flavour squarks are probed up
  to 1050, 1000, and 1325\GeV, respectively. The gluino mass is probed
  up to 1900, 1650, and 1650\GeV when the gluino decays via virtual
  states of the aforementioned squarks. The strongest mass bounds on
  the neutralinos from gluino and squark decays are 1150 and 575\GeV,
  respectively.  The search also provides sensitivity to simplified
  models inspired by split SUSY that involve the production and decay
  of long-lived gluinos. Values of the proper decay length \ctau from
  $10^{-3}$ to $10^{5}\unit{mm}$ are considered, as well as a
  metastable gluino scenario. Gluino masses up to 1750 and 900\GeV are
  probed for $\ctau = 1\unit{mm}$ and for the metastable state,
  respectively. The sensitivity is moderately dependent on model
  assumptions for $\ctau \gtrsim 1\unit{m}$. The search provides
  coverage of the \ctau parameter space for models involving
  long-lived gluinos that is complementary to existing techniques at
  the LHC.}

\hypersetup{
  pdftitle={Search for natural and split supersymmetry in
    proton-proton collisions at 13 TeV in final states with jets and
    missing transverse momentum},
  pdfauthor={CMS Collaboration},
  pdfsubject={CMS, supersymmetry, AlphaT},
  pdfkeywords={Supersymmetry, split, natural, long-lived gluinos, dark matter}
}

\maketitle

\section{Introduction}
\label{sec:introduction}

Supersymmetry (SUSY)~\cite{ref:SUSY-1, ref:SUSY0, ref:SUSY3,
  ref:SUSY1} is an extension of the standard model (SM) of particle
physics that introduces at least one bosonic (fermionic) superpartner
for each fermionic (bosonic) SM particle, where the superpartner
differs in its spin from its SM counterpart by one half unit.
Supersymmetry offers a potential solution to the hierarchy
problem~\cite{ref:hierarchy1, ref:hierarchy2}, predicts unification of
the gauge couplings at high energy~\cite{Dimopoulos:1981yj,
  Ibanez:1981yh, Marciano:1981un}, and provides a candidate for dark
matter (DM). Under the assumption of R-parity~\cite{Farrar:1978xj}
conservation, SUSY particles are expected to be produced in pairs at
the CERN LHC and to decay to the stable, lightest SUSY particle
(LSP). The LSP is assumed to be the neutralino \PSGczDo, a weakly
interacting massive particle and a viable DM
candidate~\cite{Jungman:1995df, Patrignani:2016xqp}.  So-called
natural SUSY models, which invoke only a minimal fine tuning of the
bare Higgs boson mass parameter, require only the gluino,
third-generation squarks, and a higgsino-like \PSGczDo to have masses
at or near the electroweak (EW) scale~\cite{ref:barbierinsusy}. The
interest in natural models is motivated by the discovery of a low-mass
Higgs boson~\cite{Aad:2012tfa, Chatrchyan:2012ufa, Chatrchyan:2013lba,
  Khachatryan:2014jba, Aad:2014aba, Aad:2015zhl}. The characteristic
signature of natural SUSY production at the LHC is a final state
containing an abundance of jets originating from the hadronization of
heavy-flavour quarks and significant missing transverse momentum
\ptvecmiss.

Split supersymmetry~\cite{ArkaniHamed:2004fb, Giudice:2004tc} does not
address the hierarchy problem---in contrast to natural SUSY
models---but preserves the appealing aspects of gauge coupling
unification and a DM candidate. In such a model, only the fermionic
superpartners, and a finely tuned scalar Higgs boson, may be realized
at a mass scale that is kinematically accessible at the LHC. All other
SUSY particles are assumed to be ultraheavy. Hence, within split SUSY
models, the gluino decay is suppressed because of the highly virtual
squark states. For gluino lifetimes beyond a picosecond, the gluino
hadronizes and forms a bound colour-singlet state containing the
gluino and quarks or gluons~\cite{Fairbairn:2006gg}, known as an
R-hadron, before eventually decaying to a quark--antiquark pair and
the \PSGczDo. The long-lived gluino can lead to final states with
significant \ptvecmiss from the undetected \PSGczDo particles and to
jets with vertices located a significant distance (\ie displaced) from
the luminous region of the proton beams. A metastable gluino, with a
decay length significantly beyond the scale of the CMS detector, can
escape undetected.

This paper presents a search for new-physics processes in final states
with one or more energetic jets and significant \ptvecmiss. The search
is performed with a sample of proton--proton (\Pp\Pp) collision data
at a centre-of-mass energy of 13\TeV recorded by the CMS experiment in
2016.  The analysed data sample corresponds to an integrated
luminosity of $35.9 \pm 0.9\fbinv$~\cite{CMS:2017sdi}. Earlier
searches using the same technique have been performed in {\Pp\Pp}
collisions at $\sqrt{s} = 7$, 8, and 13\TeV by the CMS
Collaboration~\cite{Khachatryan:2011tk, Chatrchyan:2011zy,
  Chatrchyan:2012wa, Chatrchyan:2013mys, Khachatryan:2016pxa,
  Khachatryan:2016dvc}. The data set analysed in this analysis is a
factor of 16 larger than that presented in
Ref.~\cite{Khachatryan:2016dvc}. The search strategy aims to provide
sensitivity to a broad range of SUSY-inspired models that predict the
existence of a DM candidate, and the search is used to constrain the
parameter spaces of a number of simplified SUSY
models~\cite{Alwall:2008ag, Alwall:2008va, sms}. The overwhelmingly
dominant background for a new-physics search in all-jet final states
resulting from {\Pp\Pp} collisions is the production of multijet events
via the strong interaction, a manifestation of quantum chromodynamics
(QCD). Several dedicated variables are employed to suppress the
multijet background to a negligible level while maintaining low
kinematical thresholds and high experimental acceptance for final
states characterized by the presence of significant \ptvecmiss. Signal
extraction is performed using additional kinematical variables, namely
the number of jets, the number of jets identified as originating from
bottom quarks, and the scalar and vector sums of the jet transverse
momenta. The ATLAS and CMS Collaborations have performed similar
searches in all-jet final states at $\sqrt{s} = 13\TeV$, of which
those providing the tightest constraints are described in
Refs.~\cite{Aaboud:2016zdn, Sirunyan:2017cwe, Sirunyan:2017kqq}. This
search does not employ specialized reconstruction
techniques~\cite{Khachatryan:2010uf, Khachatryan:2011ts, Aad:2011yf,
  Aad:2012zn, Chatrchyan:2012sp, Aad:2013gva, Khachatryan:2015jha,
  Aad:2015rba, Aaboud:2016dgf, Khachatryan:2016sfv, Aaboud:2017iio}
that target long-lived gluinos.

This paper is organized as follows. Section~\ref{sec:reconstruction}
describes the CMS apparatus and the event reconstruction
algorithms. Section~\ref{sec:selection} summarizes the selection
criteria used to identify and categorize signal events and samples of
control data. Section~\ref{sec:simulation} outlines the various
software packages used to generate the samples of simulated
events. Sections~\ref{sec:ewk} and \ref{sec:qcd} describe the methods
used to estimate the background contributions from SM processes. The
results and interpretations are described in Sections~\ref{sec:result}
and~\ref{sec:interpretations}, respectively, and summarized in
Section~\ref{sec:summary}.

\section{The CMS detector and event reconstruction}
\label{sec:reconstruction}

The central feature of the CMS detector is a superconducting solenoid
of 6\unit{m} internal diameter, providing a magnetic field of
3.8\unit{T}. Within the solenoid volume are a silicon pixel and strip
tracker, a lead tungstate crystal electromagnetic calorimeter (ECAL),
and a brass and scintillator hadron calorimeter (HCAL), each composed
of a barrel and two endcap sections. Forward calorimeters extend the
pseudorapidity coverage provided by the barrel and endcap
detectors. Muons are detected in gas-ionization chambers embedded in
the steel flux-return yoke outside the solenoid. A more detailed
description of the CMS detector, together with a definition of the
coordinate system used and the relevant kinematical variables, can be
found in Ref.~\cite{Chatrchyan:2008zzk}.

Events of interest are selected using a two-tiered trigger
system~\cite{Khachatryan:2016bia}. The first level, composed of custom
hardware processors, uses information from the calorimeters and muon
detectors to select events at a rate of around 100\unit{kHz} within a
time interval of less than 4\mus. The second level, known as the
high-level trigger, consists of a farm of processors running a version
of the full event reconstruction software optimized for fast
processing, and reduces the event rate to less than 1\unit{kHz} before
data storage. The trigger logic used by this search is summarized in
Section~\ref{sec:selection}.

\begingroup
\renewcommand*{\arraystretch}{1.2}
\newcommand{\mybox}[1]{\makebox[150pt][l]{#1}}
\begin{table}[!t]
  \topcaption{Summary of the physics object acceptances, the baseline
    event selection, the signal and control regions, and the event
    categorization schemas. The nominal categorization schema is
    defined in full in Appendix~\ref{app:suppMat}.
  }
  \label{tab:selections}
  \centering
  \resizebox{\textwidth}{!}{
    \begin{tabular}{ ll }
      \hline
      \multicolumn{2}{l}{\textbf{Physics object acceptances}}                                                                            \\
      Jet                               & $\pt > 40\GeV$, $\abs{\eta} < 2.4$                                                             \\
      Photon                            & $\pt > 25\GeV$, $\abs{\eta} < 2.5$, isolated in cone ${\Delta}R < 0.3$                         \\
      Electron                          & $\pt > 10\GeV$, $\abs{\eta} < 2.5$, $I^\text{rel} < 0.1$ in cone $0.05 < {\Delta}R(\pt) < 0.2$ \\
      Muon                              & $\pt > 10\GeV$, $\abs{\eta} < 2.5$, $I^\text{rel} < 0.2$ in cone $0.05 < {\Delta}R(\pt) < 0.2$ \\
      Single isolated track (SIT)       & $\pt > 10\GeV$, $\abs{\eta} < 2.5$, $I^\text{track} < 0.1$ in cone ${\Delta}R < 0.3$           \\
      \hline
      \multicolumn{2}{l}{\textbf{Baseline event selection}}                                                                              \\
      All-jet final state               & Veto events containing photons, electrons, muons, and SITs within acceptance                   \\
      \ptmiss quality                   & Veto events based on filters related to beam and instrumental effects                          \\
      Jet quality                       & Veto events containing jets that fail identification criteria or $0.1 < \fhleadjet < 0.95$     \\
      Jet energy and sums               & $\pt^{\jet{1}} > 100\GeV$, $\scalht > 200\GeV$, $\mht > 200\GeV$                               \\
      Jets outside acceptance           & $\mhtmet < 1.25$, veto events containing jets with $\pt > 40\GeV$ and $\abs{\eta} > 2.4$       \\
      \hline
      \textbf{Signal region}            & Baseline selection +                                                                           \\
      \alphat threshold (\scalht range) & 0.65 (200--250\GeV), 0.60 (250--300), 0.55 (300--350), 0.53 (350--400), 0.52 (400--900)        \\
      \bdphi threshold                  & $\bdphi > 0.5$ ($\njet \geq 2$), $\bdphimod > 0.5$ ($\njet = 1$)                               \\
      \hline
      \multicolumn{2}{l}{\textbf{Nominal categorization schema}}                                                                         \\
      \njet                             & \mybox{1} (monojet)                                                                            \\
                                        & \mybox{${\geq}2a$} ($a$ denotes asymmetric, $40 < \pt^{\jet{2}} < 100\GeV$)                    \\
                                        & \mybox{2, 3, 4, 5, ${\geq}6$} (symmetric, $\pt^{\jet{2}} > 100\GeV$)                           \\
      \nb                               & \mybox{0, 1, 2, 3, ${\geq}4$} (can be dropped/merged \vs \njet)                                \\
      \scalht boundaries                & \mybox{200, 400, 600, 900, 1200\GeV} (can be dropped/merged \vs \njet, \nb)                    \\
      \mht boundaries                   & \mybox{200, 400, 600, 900\GeV} (can be dropped/merged \vs \njet, \nb, \scalht)                 \\
      \hline
      \multicolumn{2}{l}{\textbf{Simplified categorization schema}}                                                                      \\
      Topology (\njet, \nb)
                                        & \mybox{Monojet-like} ($1 \cap {\geq}2a, 0$), ($1 \cap {\geq}2a, {\geq}1$)                      \\
                                        & \mybox{Low \njet} ($2 \cap 3, 0 \cap 1$), ($2 \cap 3, {\geq}2$)                                \\
                                        & \mybox{Medium \njet} ($4 \cap 5, 0 \cap 1$), ($4 \cap 5, {\geq}2$)                             \\
                                        & \mybox{High \njet} (${\geq}6, 0 \cap 1$), (${\geq}6, {\geq}2$)                                 \\
      \scalht boundaries                & $\scalht > 200\GeV$ ($\njet \leq 3$), $\scalht > 400\GeV$ ($\njet \geq 4$)                     \\
      \mht boundaries                   & 200, 400, 600, 900\GeV                                                                         \\
      \hline
      \textbf{Control regions}          & Baseline selection +                                                                           \\
      \mj (inverted $\mu$ veto)
                                        & $\pt^{\mu_1} > 30\GeV$, $\abs{\eta^{\mu_1}} < 2.1$,
                                        ${\Delta}R(\mu,\jet{i}) > 0.5$,
                                        $30 < m_\mathrm{T}(\ptvec^\mu,\ptvecmiss) < 125\GeV$                                             \\
      \mmj (inverted $\mu$ veto)
                                        & $\pt^{\mu_{1,2}} > 30\GeV$, $\abs{\eta^{\mu_{1,2}}} < 2.1$,
                                        ${\Delta}R(\mu_{1,2},\jet{i}) > 0.5$,
                                        $ \abs{m_{\mu\mu} - m_\mathrm{\cPZ}} < 25\GeV$                                                   \\
      Multijet-enriched                 & Sidebands to signal region: $\mht/\ptmiss > 1.25$ and/or $\bdphi < 0.5$                        \\
      \hline
    \end{tabular}
  }
\end{table}
\endgroup

The particle-flow (PF) event algorithm~\cite{CMS-PRF-14-001}
reconstructs and identifies each individual particle with an optimized
combination of information from the various elements of the CMS
detector. In this process, the identification of the particle type
(photon, electron, muon, charged hadron, neutral hadron) plays an
important role in the determination of the particle direction and
energy. The energy of photons~\cite{Khachatryan:2015iwa} is directly
obtained from the ECAL measurement, corrected for zero-suppression
effects. The energy of electrons~\cite{Khachatryan:2015hwa} is
determined from a combination of the electron momentum at the primary
interaction vertex as determined by the tracker, the energy of the
corresponding ECAL measurement, and the energy sum of all
bremsstrahlung photons spatially compatible with originating from the
electron track. The energy of muons~\cite{Chatrchyan:2012xi} is
obtained from the curvature of the corresponding track. The energy of
charged hadrons is determined from a combination of their momentum
measured in the tracker and the matching ECAL and HCAL energy
deposits, corrected for zero-suppression effects and for the response
function of the calorimeters to hadronic showers. Finally, the energy
of neutral hadrons is obtained from the corresponding corrected ECAL
and HCAL energy. The reconstruction techniques used by this search are
not specialized to target specific experimental signatures (such as
displaced jets). The physics objects used in this search are defined
below and are summarized in Table~\ref{tab:selections}. In the case of
photons and leptons, further details can be found in
Ref.~\cite{Khachatryan:2016dvc} and references therein.

The reconstructed vertex with the largest value of summed physics
object $\pt^2$ is taken to be the primary {\Pp\Pp} interaction vertex
(PV). The physics objects considered are those returned by a jet
finding algorithm~\cite{Cacciari:2008gp, Cacciari:2011ma} applied to
all charged particle tracks associated with the vertex, and the
associated \ptmiss, taken as the negative vector sum of the \pt of
those physics objects. Charged particle tracks associated with
vertices from additional {\Pp\Pp} interactions within the same or
nearby bunch crossings (pileup) are not considered by the PF algorithm
as part of the global event reconstruction. The energy deposit
associated with each physics object %PF particle candidates
is corrected to account for contributions from neutral particles
originating from pileup interactions~\cite{Cacciari:2007fd}.

Samples of signal events and control data are defined, respectively,
by the absence or presence of photons and leptons that are isolated
from other activity in the event. Photons are required to be
isolated~\cite{Khachatryan:2015iwa} within a cone around the photon
trajectory defined by the radius ${\Delta}R =
\sqrt{\smash[b]{(\Delta\phi)^2 + (\Delta\eta)^2}} = 0.3$, where
$\Delta\phi$ and $\Delta\eta$ represent differences in the azimuthal
angle (radians) and pseudorapidity. Isolation for an electron or muon
is a relative quantity, $I^\text{rel}$, defined as the scalar \pt sum
of all PF particle candidates within a cone around its trajectory,
divided by the lepton \pt. The cone radius is dependent on the lepton
\pt, ${\Delta}R = \min [ \max( 0.05, 10\GeV / \pt ), 0.2 ]$, to
maintain high efficiency for semileptonic decays of Lorentz-boosted
top quarks~\cite{Rehermann:2010vq}. Isolated electrons and muons are
required to satisfy $I^\text{rel} < 0.1$ and 0.2, respectively.
Electron and muon candidates that fail any of the aforementioned
requirements, as well as charged-hadron candidates from hadronically
decaying tau leptons, are collectively labelled as single isolated
tracks (SITs) if the scalar \pt sum of additional tracks associated
with the PV within a cone ${\Delta}R < 0.3$ around the track
trajectory, relative to the track \pt, satisfies $I^\text{track} <
0.1$. All isolation variables exclude the contributions from the
physics object itself and pileup
interactions~\cite{Khachatryan:2015iwa, Khachatryan:2015hwa,
  Chatrchyan:2012xi}. The experimental acceptances for photons,
electrons, muons, and SITs are defined by the transverse momentum
requirements $\pt > 25$, 10, 10, and 10\GeV, respectively, and the
pseudorapidity requirement $\abs{\eta} < 2.5$.

Jets are reconstructed from the PF particle candidates, clustered by
the anti-\kt algorithm~\cite{Cacciari:2008gp, Cacciari:2011ma} with a
distance parameter of 0.4. In this process, the raw jet energy is
obtained from the sum of the particle candidate energies, and the raw
jet momentum by the vectorial sum of the particle candidate momenta,
which results in a nonzero jet mass. An offset correction is applied
to jet energies to take into account the contributions from neutral
particles produced in pileup interactions~\cite{Cacciari:2007fd,
  CMS-PAS-JME-14-001}. The raw jet energies are then corrected to
establish a relative uniform response of the calorimeter in $\eta$ and
a calibrated absolute response in \pt. Jet energy corrections are
derived from simulation, and are confirmed with in situ measurements
of the energy balance in dijet, multijet, \gj, and leptonically
decaying \zj events~\cite{Khachatryan:2016kdb}. Jets are required to
satisfy $\pt > 40\GeV$ and $\abs{\eta} < 2.4$. Jets are also subjected
to a standard set of identification criteria~\cite{2011JInst611002C}
that require each jet to contain at least two particle candidates and
at least one charged particle track, and the energy fraction \fh
attributed to charged-hadron particle candidates is required to be
nonzero.

Jets can be identified as originating from b quarks using the combined
secondary vertex (CSVv2) algorithm~\cite{BTV-16-002}.  Data samples
are used to measure the b tagging efficiency, which is the probability
to correctly identify jets originating from b quarks, as well as the
mistag probabilities for jets that originate from light-flavour (LF)
partons (u, d, s quarks or gluon) or a charm quark. A working point is
employed that yields a b tagging efficiency of ${\approx}69\%$ for
jets with $\pt > 30\GeV$ from \ttbar events, and charm and LF mistag
probabilities of ${\approx}18$ and ${\approx}1\%$, respectively, for
multijet events.

Finally, the most accurate estimator for \ptvecmiss is defined as the
projection on the plane perpendicular to the beams of the negative
vector sum of the momenta of all PF particle candidates in an
event. Its magnitude is referred to as \ptmiss.

\section{Event selection and categorization}
\label{sec:selection}

A baseline set of event selection criteria, described in
Section~\ref{sec:baseline}, is used as a basis for all data samples
used in this search. Two additional requirements, described in
Section~\ref{sec:signal}, are employed to define a sample of signal
events, labelled henceforth as the signal region (SR). The
categorization of signal events and the background composition are
described in Sections~\ref{sec:categorization} and \ref{sec:bkgd},
respectively. Three independent control regions (CRs), comprising
large samples of event data, are defined by the selection criteria
described in Section~\ref{sec:control}. All selection criteria are
summarized in Table~\ref{tab:selections}.

\subsection{Baseline selections}
\label{sec:baseline}

Events containing isolated photons, electrons and muons, or SITs that
satisfy the requirements summarized in Table~\ref{tab:selections} are
vetoed. The aforementioned vetoes are employed to select all-jet final
states, suppress SM processes that produce final states containing
neutrinos, and reduce backgrounds from misreconstructed or nonisolated
leptons as well as single-prong hadronic decays of $\tau$ leptons.

Beam halo, spurious jet-like features originating from isolated noise
patterns in the calorimeter systems, detector inefficiencies, and
reconstruction failures can all lead to large values of \ptmiss. Such
events are rejected with high efficiency using dedicated
vetoes~\cite{CMS-PAS-JME-16-004, Khachatryan:2014gga}. Events are
vetoed if any jet fails the identification criteria described in
Section~\ref{sec:reconstruction}. Further, \fh for the highest \pt jet
of the event, $\jet{1}$, is required to satisfy $0.1 < \fhleadjet <
0.95$ to further suppress beam halo and rare reconstruction failures.

The highest \pt jet in the event is required to satisfy $\pt^{\jet{1}}
> 100\GeV$. The mass scale of each event is estimated from the scalar
\pt sum of the jets, defined as $\scalht = \sum_{\jet{i} = 1}^{\njet}
\pt^{\,\jet{i}}$, where \njet is the number of jets within the
experimental acceptance. The estimator for \ptvecmiss used by this
search is given by the magnitude of the vector \pt sum of the jets,
$\mht = |\sum_{\jet{i} = 1}^{\njet} \ptvec^{\,\jet{i}}|$. Significant
hadronic activity and \ptvecmiss, typical of SUSY processes, is
ensured by requiring $\scalht > 200\GeV$ and $\mht > 200\GeV$,
respectively.

Events are vetoed if any additional jet satisfies $\pt > 40\GeV$ and
$|\eta| > 2.4$ to maintain the resolution of the \mht variable.  An
additional veto is employed to deal with the circumstance in which
several jets with transverse momentum below the \pt thresholds and
collinear in $\phi$ can result in significant \mht relative to
\ptmiss, the latter of which is less sensitive to jet thresholds. This
type of event topology, which is typical of multijet events, is
suppressed while maintaining high efficiency for new-physics processes
with significant \ptvecmiss by requiring $\mhtmet < 1.25$.

\subsection{Signal region}
\label{sec:signal}

The multijet background dominates over all other SM backgrounds
following the application of the baseline event selection
criteria. The multijet background is suppressed to a negligible level
through the application of two dedicated variables that provide strong
discrimination between multijet events with \ptvecmiss resulting from
instrumental sources, such as jet energy mismeasurements, and
new-physics processes that involve the production of weakly
interacting particles that escape detection.

The first variable, \alphat~\cite{Randall:2008rw, Khachatryan:2011tk},
is designed to be intrinsically robust against jet energy
mismeasurements. In its simplest form, the \alphat variable is defined
as $\alphat = \ET^{\jet{2}}/M_\mathrm{T}$, where $M_\mathrm{T} = \sqrt{\smash[b]{ 2
  \ET^{\jet{1}} \ET^{\jet{2}} (1 - \cos\phi_{\jet{1},\jet{2}})}}$ and
$\phi_{\jet{1},\jet{2}}$ is defined as the azimuthal angle between
jets $\jet{1}$ and $\jet{2}$. In the absence of jet energy
mismeasurements, and in the limit for which the \ET of each jet is
large compared with its mass, a well-measured dijet event with
$\ET^{\jet{1}} = \ET^{\jet{2}}$ and back-to-back jets
($\phi_{\jet{1},\jet{2}} = \pi$) yields an \alphat value of 0.5. In
the presence of a jet energy mismeasurement, $\ET^{\jet{1}} >
\ET^{\jet{2}}$ and $\alphat < 0.5$. Values significantly greater than
0.5 can be observed when the two jets are not back-to-back and recoil
against \ptvecmiss from weakly interacting particles that escape the
detector. The definition of the \alphat variable can be generalized
for events with two or more jets, as described in
Ref.~\cite{Khachatryan:2011tk}. Multijet events populate the region
$\alphat \lesssim 0.5$ and the \alphat distribution is characterized
by a sharp edge at 0.5, beyond which the multijet event yield falls by
several orders of magnitude. The SM backgrounds that involve the
production of neutrinos result in a long tail in \alphat beyond values
of 0.5. A \scalht-dependent \alphat threshold that decreases from 0.65
at low \scalht to 0.52 at high \scalht within the range $200 < \scalht
< 900\GeV$ is employed to maintain an approximately constant rejection
power against the multijet background.

The second variable, known as \bdphi, considers the minimum azimuthal
angular separation between each jet in the event and the vector \pt
sum of all other jets in the event. Multijet events typically populate
the region $\bdphi \approx 0$ while events with genuine \ptvecmiss can
have values up to $\bdphi = \pi$. The requirement $\bdphi > 0.5$ is
sufficient to reduce significantly the multijet background, including
rare contributions from energetic multijet events that yield both high
jet multiplicities and significant \ptvecmiss because of
high-multiplicity neutrino production in semileptonic heavy-flavour
decays. For events that satisfy $\njet = 1$, a small modification to
the \bdphi variable is utilized that considers any additional jets
with $25 < \pt < 40\GeV$ that are outside the nominal experimental
acceptance ($\bdphimod > 0.5$).

The requirements on \alphat and \bdphi, summarized in
Table~\ref{tab:selections}, suppress the expected contribution from
multijet events to the sub-percent level with respect to the total
expected background counts from all other SM processes. For the region
$\scalht > 900\GeV$, the necessary control of the multijet background
is achieved solely with the \bdphi and \bdphimod variables. These
requirements complete the definition of the SR.

Signal events are recorded with a number of trigger algorithms. Events
with $\njet \geq 2$ must satisfy thresholds on both \scalht and
\alphat that are looser than those used to define the
SR. High-activity events that satisfy $\scalht > 900\GeV$ are also
recorded. Finally, a trigger condition that requires $\mht > 120\GeV$,
$\ptmiss > 120\GeV$, and a single jet with $\pt > 20\GeV$ and $|\eta|
< 5.2$ is also used to efficiently record signal events for all
categories of the SR, including those that satisfy $\njet \geq 1$. The
combined performance of these trigger algorithms yields high
efficiencies, as determined from samples of CR data enriched in vector
boson + jets and \ttbar events. The efficiencies are primarily
\scalht-dependent and range from 97.4--97.9\% ($200 < \scalht <
600\GeV$) to 100\% ($\scalht > 600\GeV$) with statistical and
systematic uncertainties at the percent level. Trigger efficiencies
for a range of benchmark signal models are typically comparable or
higher (${\approx}100\%$).

\subsection{Event categorization}
\label{sec:categorization}

Signal events are categorized into 27 discrete topologies according to
\njet and the number of b-tagged jets \nb. Events are further binned
according to the energy sums \scalht and \mht. The binning schema is
determined primarily by the statistical power of the \mj and \mmj CRs.

Seven bins in \njet are considered, as summarized in
Table~\ref{tab:selections}. Events that contain only a single jet
within the experimental acceptance ($\njet = 1$) are labelled as
``monojet''. Events containing two or more jets are categorized
according to the second-highest jet \pt. Events that satisfy $\njet
\geq 2$ with only the highest \pt jet satisfying $\pt > 100\GeV$ are
labelled as ``asymmetric''. Events for which the second-highest jet
\pt also satisfies $\pt > 100\GeV$ are labelled as ``symmetric'' and
are categorized according to \njet (2, 3, 4, 5, and ${\geq}6$). The
symmetric topology targets the pair production of SUSY particles and
their prompt cascade decays, while the monojet and asymmetric
topologies preferentially target models with a compressed mass
spectrum and long-lived SUSY particles.

Events are also categorized according to \nb (0, 1, 2, 3, ${\geq}4$),
where \nb is bounded from above by \njet and the choice of
categorization is dependent on \njet. Higher \nb multiplicities target
the production of third-generation squarks.

The nominal binning schema for \scalht is defined as follows: four
bounded bins that satisfy 200--400, 400--600, 600--900, and
900--1200\GeV, and a final open bin $\scalht > 1200\GeV$. This schema
is adapted per (\njet, \nb) category as follows: only the region
$\scalht > 400\GeV$ is considered for events that satisfy $\njet \geq
4$, and bins at high \scalht are merged with lower-\scalht bins to
satisfy a threshold of at least four events in the corresponding bins
of the CRs.

The \mht variable is used to further categorize events according to
three bounded bins that satisfy 200--400, 400--600, and 600--900, and
a final open bin $\mht > 900\GeV$. The \mht binning depends on \njet,
\nb, and \scalht. Given that \mht cannot exceed \scalht by
construction, the lower bound of the final \mht bin is restricted to
be not higher than the lower bound of the \scalht bin in
question. Events that satisfy $\njet = 1$ or $200 < \scalht < 400\GeV$
are not categorized according to \mht.

In total, there are 254 bins in the SR. An alternate, simplified
binning schema is also provided in which events are categorized
according to eight topologies defined in terms of \njet and \nb. For
each topology, event yields are integrated over the full available
\scalht range and categorized according to the four nominal \mht bins
defined above. This schema has 32 bins that are exclusive, contiguous,
and provide a complete coverage of the SR. The SM background estimates
are obtained from the same likelihood model as the one used to
determine the nominal result.

\subsection{Background composition}
\label{sec:bkgd}

Following the application of the SR selection criteria, the multijet
background is reduced to a negligible level. The remaining background
contributions are dominated by processes that involve the production
of high-\pt neutrinos in the final state. The associated production of
jets and a \cPZ\ boson that decays to $\cPgn\cPagn$ dominates the
background contributions for events containing low numbers of jets and
b-tagged jets. The \znunuj background is irreducible. The associated
production of jets and a \PW\ boson that decays to $\ell\nu$
($\ell=\Pe$, $\Pgm$, $\Pgt$) is also a significant background in the
same phase space. The production and semileptonic decay of top
quark-antiquark pairs (\ttbar) becomes the dominant background process
for events containing high numbers of jets or b-tagged jets. Events
that contain the leptonic decay of a \PW\ boson are typically rejected
by the vetoes that identify the presence of leptons or single isolated
tracks. If the lepton is outside the experimental acceptance, or is
not identified or isolated, then the event is not vetoed and the
aforementioned processes lead to what is collectively known as the
``lost lepton'' (\lost) background. Residual contributions from other
SM processes are also considered, such as single top quark production;
\PW\PW, \PW\cPZ, \cPZ\cPZ\ (diboson) production; and the associated
production of \ttbar and a boson ($\ttbar\PW$, $\ttbar\cPZ$,
$\ttbar\gamma$, and $\ttbar\PH$).

\subsection{Control regions}
\label{sec:control}

Topological and kinematical requirements, summarized in
Table~\ref{tab:selections}, ensure that the samples of CR data are
enriched in the same or similar SM processes that populate the SR, as
well as being depleted in contributions from SUSY processes (signal
contamination).

Three sidebands to the SR comprising multijet-enriched event samples
are defined by: $1.25 < \mhtmet < 3.0$ (region $A$), $0.2 < \bdphi <
0.5$ ($B$), and both $1.25 < \mhtmet < 3.0$ and $0.2 < \bdphi < 0.5$
($C$). Events are categorized according to \njet and \scalht,
identically to the SR. Events are recorded with the signal triggers
described above.

Two additional CRs comprising \mj and \mmj event samples are defined
by the application of the baseline selections and requirements on
isolated, central, high-\pt muons. Tighter isolation requirements for
the muons are applied with respect to those indicated in
Table~\ref{tab:selections}.  A trigger condition that requires an
isolated muon with $\pt > 24\GeV$ and $\abs{\eta} < 2.1$ is used to
record the \mj and \mmj event samples with efficiencies of
${\approx}90$ and ${\approx}99\%$, respectively. For both samples, no
requirements on \alphat or \bdphi are imposed. The kinematical
properties of events in the \mj and \mmj CRs and SR are comparable
once the muon or dimuon system is ignored in the calculation of
event-level quantities such as \scalht and \mht.  Events in both
samples are categorized according to \njet, \scalht, and \nb, with
counts integrated over \mht. The \njet categorization is identical to
the SR. Background predictions are determined using up to eleven bins
in \scalht that are then aggregated to match the \scalht binning
schema used by the SR. The \nb categorization for \mj events is
identical to the SR, whereas \mmj events are subdivided according to
$\nb = 0$ and $\nb \geq 1$. Differences in the binning schemas between
the SR and CRs are accounted for in the background estimation methods
through simulation-based templates, the modelling of which is
validated against control data.

The \mj event sample is enriched in events from \wmj and \ttbar
production, as well as other SM processes (\eg single top quark and
diboson production), that are manifest in the SR as the \lost
backgrounds. Each event is required to contain a single isolated muon
with $\pt > 30\GeV$ and $\abs{\eta} < 2.1$ to satisfy trigger
conditions, and is well separated from each jet $\mathrm{j}_i$ in the
event according to ${\Delta}R(\mu,\mathrm{j}_i) > 0.5$. The transverse
mass $m_\mathrm{T} = \sqrt{\smash[b]{2\pt^\mu\ptmiss[1-
    \cos(\Delta\phi_{\mu,\ptvecmiss})]}}$, where
$\Delta\phi_{\mu,\ptvecmiss}$ is the difference between the azimuthal
angles of the muon transverse momentum vector $\ptvec^\mu$ and of
$\ptvecmiss$, must fall within the range 30--125\GeV to select a
sample of events rich in \PW\ bosons.

The \mmj sample is enriched in $\cPZ\! \to\!  \mu^+\mu^-$ events that
have similar acceptance and kinematical properties to \znunuj events
when the muons are ignored. The sample uses selection criteria similar
to the \mj sample, but requires two oppositely charged, isolated muons
that both satisfy $\pt > 30\GeV$, $\abs{\eta} < 2.1$, and
${\Delta}R(\mu_{1,2},\mathrm{j}_i) > 0.5$. The muons are also required
to have a dilepton invariant mass $m_{\mu\mu}$ within a ${\pm}25\GeV$
window around the mass of the \cPZ\ boson~\cite{Patrignani:2016xqp}.

\section{Monte Carlo simulation}
\label{sec:simulation}

The search relies on several samples of simulated events, produced
with Monte Carlo (MC) generator programs, to aid the estimation of SM
backgrounds and evaluate potential signal contributions.

The \MGvATNLO 2.2.2~\cite{Alwall2014} event generator is used at
leading-order (LO) accuracy to produce samples of \wj, \zj, \ttbar,
and multijet events. Up to three or four additional partons are
included in the matrix-element calculation for \ttbar and vector boson
production, respectively. Simulated \wj and \zj events are weighted
according to the true vector boson \pt to account for the effect of
missing next-to-leading-order (NLO) QCD and EW terms in the
matrix-element calculation~\cite{Alwall2014, Kuhn:2005gv}, according
to the method described in Ref.~\cite{Khachatryan:2016mdm}. Within the
range of vector boson \pt that can be probed by this search, the QCD
and EW corrections~\cite{Kuhn:2005gv} are largest, ${\approx}40\%$ and
${\approx}15\%$, at low and high values of boson \pt,
respectively. Simulated \ttbar events are weighted to improve the
description of jets arising from initial-state radiation
(ISR)~\cite{Chatrchyan:2013xna}. The weights vary from 0.92 to 0.51
depending on the number of jets (1--6) from ISR, with an uncertainty
of one half the deviation from unity. The \MGvATNLO generator is used
at NLO accuracy to generate samples of $s$-channel production of
single top quark, as well as $\ttbar\PW$ and $\ttbar\cPZ$ events. The
NLO \POWHEG v2~\cite{powheg, powheg_top_Wt} generator is used to
describe the $t$- and \PW\cPqt-channel production of events containing
single top quarks, as well as $\ttbar\PH$ events. The \PYTHIA
8.205~\cite{pythia} program is used to generate diboson (\PW\PW,
\PW\cPZ, \cPZ\cPZ) production.

Event samples for signal models involving the production of gluino or
squark pairs, in association with up to two additional partons, are
generated at LO with \MGvATNLO, and the decay of the SUSY particles is
performed with the \PYTHIA program. The {NNPDF}3.0 LO and
{NNPDF}3.0 NLO~\cite{nnpdf} parton distribution functions
(PDFs) are used, respectively, with the LO and NLO generators
described above.

The simulated samples for SM processes are normalized according to
production cross sections that are calculated with NLO and next-to-NLO
precision~\cite{Alwall2014, wphys, fewz, wwxs, top++, nlotop,
  powheg_top_Wt}. The production cross sections for pairs of gluinos
or squarks are determined at NLO plus next-to-leading-logarithm (NLL)
precision~\cite{Beenakker:1996ch, Kulesza:2008jb, Kulesza:2009kq,
  Beenakker:2009ha, Beenakker:2011fu, Borschensky:2014cia}. All other
SUSY particles, apart from the \PSGczDo, are assumed to be heavy and
decoupled from the interaction. Uncertainties in the cross sections
are determined from different choices of PDF sets, and factorization
and renormalization scales ($\mu_\mathrm{F}$ and $\mu_\mathrm{R}$),
according to the prescription in Ref.~\cite{Borschensky:2014cia}. The
\PYTHIA program with the \textsc{CUETP8M1} tune~\cite{Skands:2014pea,
  Khachatryan:2015pea} is used to describe parton showering and
hadronization for all simulated samples.

The \textsc{rhadrons} package within the \PYTHIA 8.205 program is used
to describe the formation of R-hadrons through the hadronization of
gluinos~\cite{Fairbairn:2006gg, Kraan:2004tz, Mackeprang:2006gx}. The
hadronization process, steered according to the default parameter
settings of the \textsc{rhadrons} package, predominantly yields
meson-like (\PSg\Pq\Paq) and baryon-like (\PSg\Pq\Pq\Pq) states, as
well as glueball-like (\PSg\Pg) states with a probability $P_{\PSg\Pg}
= 10\%$, where \PSg, \Pg, \Pq, and \Paq\ represent a gluino, gluon,
quark, and antiquark, respectively. The gluino is assumed to undergo a
three-body decay, to a \Pq\Paq\ pair and the \PSGczDo, according to
its proper decay length \ctau that is a parameter of the simplified
model~\cite{Buchmueller:2017uqu}. Studies with alternative values for
parameters that influence the hadronization of the gluino, such as
$P_{\PSg\Pg} = 50\%$, indicate a minimal influence on the event
topology and kinematical variables for the models considered in this
paper. Further, the model-dependent interactions of R-hadrons with the
detector material are not considered by default, as studies
demonstrate that the sensitivity of this search is only moderately
dependent on these interactions, as discussed in
Section~\ref{sec:interpretations}.

The description of the detector response is implemented using the
\GEANTfour~\cite{geant} package for all simulated SM processes. Scale
factors are applied to simulated event samples that correct for
differences with respect to data in the b tagging efficiency and
mistag probabilities. The scale factors have typical values of
${\approx}$0.95--1.00 and ${\approx}$1.00--1.20, respectively, for a
jet \pt range of 40--600\GeV~\cite{BTV-16-002}. All remaining signal
models rely on the CMS fast simulation package~\cite{fastsim} that
provides a description that is consistent with \GEANTfour following
the application of near-unity corrections for the differences in the b
tagging efficiency and mistag probabilities, as well as corrections
for the differences in the modelling of the \mht distribution. To
model the effects of pileup, all simulated events are generated with a
nominal distribution of {\Pp\Pp} interactions per bunch crossing and then
weighted to match the pileup distribution as measured in data.

\section{Nonmultijet background estimation}
\label{sec:ewk}

The \lost and \znunuj backgrounds, collectively labelled henceforth as
the nonmultijet backgrounds, are estimated from data samples in CRs
and transfer factors $\mathcal{R}$ determined from the ratios of
expected counts obtained from simulation:
\begin{align}
  \tf^{\lost} & =
  \frac{N^{\lost}_\text{MC}(\njet, \scalht, \nb, \mht)}
  {N^{\mj}_\text{MC}(\njet, \scalht, \nb)\hfill},
  &
  N^{\lost}_\text{pred}  & =
  \tf^{\lost} \; N^{\mj}_\text{data},
  \\
  \tf^{\znunu}  & = 
  \frac{N^{\znunu}_\text{MC}(\njet, \scalht, \nb, \mht)}
  {N^{\mmj}_\text{MC}(\njet, \scalht, \nb)\hfill},
  &
  N^\text{\znunu}_\text{pred}  & = 
  \tf^{\znunu} \; N^{\mmj}_\text{data},
\end{align}
where $\tf^{\lost}$ and $\tf^{\znunu}$ are the transfer factors that
act as multiplier terms on the event counts $N^{\mj}_\text{data}$ and
$N^{\mmj}_\text{data}$ observed in each (\njet, \scalht, \nb) bin of,
respectively, the \mj and \mmj CRs to estimate the \lost or \znunuj
background counts $N^{\lost}_\text{pred}$ and $N^{\znunu}_\text{pred}$
in the corresponding (\njet, \scalht, \nb, \mht) bins of the
SR. Several sources of uncertainty in the transfer factors are
evaluated.  In addition to statistical uncertainties arising from
finite-size simulated event samples, the most relevant systematic
effects are discussed below.

The uncertainties from known theoretical and experimental sources are
propagated through to the transfer factors to ascertain the magnitude
of variations related to the following: the jet energy scale, the
scale factors related to the b tagging efficiency and mistag
probabilities, the efficiency to trigger on and identify, or veto,
well-reconstructed isolated leptons, the
PDFs~\cite{Butterworth:2015oua}, $\mu_\mathrm{F}$ and $\mu_\mathrm{R}$,
and the modelling of jets from ISR produced in association with
\ttbar~\cite{Chatrchyan:2013xna}. Uncertainties of 100\% in both the
NLO QCD and EW corrections to the \wj and \zj simulated samples are
also considered. A 5\% uncertainty in the total inelastic cross
section~\cite{Aaboud:2016mmw} is assumed and propagated through to the
weighting procedure to account for differences between the data and
simulation in the pileup distributions. Uncertainties in the signal
trigger efficiency measurements are also propagated to the transfer
factors. The effects of the aforementioned systematic uncertainties
are summarized in Table~\ref{tab:bkgd_systs}, in terms of
representative ranges.  Each source of uncertainty is assumed to vary
with a fully correlated behaviour across the full phase space of the
SR and CRs.

\begingroup
\renewcommand*{\arraystretch}{1.2}
\begin{table}[!t]
  \topcaption{
    Systematic uncertainties in the $\lost$ and $\znunu$ background
    evaluation. The quoted ranges are representative of the minimum
    and maximum variations observed across all bins of the signal
    region. Pairs of ranges are quoted for uncertainties determined
    from closure tests in data, which correspond to variations as a
    function of \njet and \scalht, respectively.
  }
  \label{tab:bkgd_systs}
  \centering
  \begin{tabular}{ lcc }
    \hline
    Source of uncertainty                        & \multicolumn{2}{c}{Magnitude [\%]} \\
    \cline{2-3}
                                                 & $\lost$      & $\znunu$            \\
    \hline
    Finite-size simulated samples                & 1--50        & 1--50               \\
    Total inelastic cross section (pileup)       & 0.6--3.8     & 2.3--2.8            \\
    $\mu_\mathrm{F}$ and $\mu_\mathrm{R}$ scales & 2.3--3.6     & 0.9--4.7            \\
    Parton distribution functions                & 1.1--2.7     & 0.0--3.3            \\
    \wj cross section                            & 0.2--1.4     & \NA                 \\
    \ttbar cross section                         & 0.0--1.0     & \NA                 \\
    NLO QCD corrections                          & 1.5--13      & 2.6--17             \\
    NLO EW corrections                           & 0.1--9.5     & 0.0--7.8            \\
    ISR (\ttbar)                                 & 0.8--1.1     & \NA                 \\
    Signal trigger efficiency                    & 0.0--3.1     & 0.0--2.0            \\
    Lepton efficiency (selection)                & 2.0          & 4.0                 \\
    Lepton efficiency (veto)                     & 5.0          & 5.0                 \\
    Jet energy scale                             & 3.4--5.5     & 5.3--8.0            \\
    b tagging efficiency                         & 0.4--0.6     & 0.3--0.6            \\
    Mistag probabilities                         & 0.1--1.4     & 0.2--1.8            \\
    \alphat extrapolation                        & 3--9, 2--6   & 3--9, 2--6          \\
    \bdphi extrapolation                         & 3--22, 2--18 & 3--22, 2--18        \\
    \PW\ boson polarization                      & 1--7, 2--7   & \NA                 \\
    Single isolated track veto                   & 0--10, 0--13 & \NA                 \\
    \hline
  \end{tabular}
\end{table}
\endgroup

Sources of additional uncertainties are determined from closure tests
performed using control data that aim to identify \njet- or
\scalht-dependent sources of systematic bias arising from
extrapolations in kinematical variables covered by the transfer
factors. Several sets of tests are performed. The accuracy of the
modelling of the efficiencies of both the \alphat and \bdphi
requirements is estimated from both the \mj and \mmj samples. The
effects of \PW\ boson polarization are probed by using \mj events with
a positively charged muon to predict those containing a negatively
charged muon. Finally, the efficiency of the single isolated track
veto is also probed using a sample of \mj events. The uncertainties
are summarized in Table~\ref{tab:bkgd_systs}.

The simulation modelling of the \nb distributions for the \znunuj
background in the region $\nb \geq 1$ is evaluated through a binned
maximum-likelihood fit to the observed \nb distributions in data in
each (\njet, \scalht) bin of the \mmj CR. Additional checks are
performed in \mmj samples that are enriched in mistagged jets that
originate from LF partons or charm quarks, or the genuine tags of b
quarks from gluon splitting, through the use of loose and tight
working points of the b tagging algorithm, respectively. No tests
reveal evidence of significant bias in the simulation modelling of the
\nb distribution.

Finally, the modelling of the \mht distribution in simulated events is
compared to the distributions observed in \mj and \mmj control data,
and inspected for trends, by assuming a linear behaviour of the ratio
of observed and simulated counts as a function of \mht. Linear fits
are performed independently for each \njet category while integrating
event counts over \nb and \scalht, and then repeated for each \scalht
bin while integrating event counts over \njet and \nb. Systematic
uncertainties are determined from any nonclosure between data and
simulation as a function of \njet and are assumed to be correlated in
\scalht (and \nb), and vice versa. The uncertainties can be as large
as ${\approx}50\%$ in the most sensitive \mht bins.

\section{Multijet background estimation}
\label{sec:qcd}

The multijet background is estimated using the three data sidebands
defined in Section~\ref{sec:control}. Events in each sideband are
categorized according to \njet and \scalht. The event counts in data
are corrected to account for contamination from nonmultijet SM
processes, such as vector boson and \ttbar production, as well as the
residual contributions from other SM processes. The nonmultijet
processes are estimated from the \mj and \mmj CRs, following a
procedure similar to the one described in Section~\ref{sec:ewk}. The
corrected counts are assumed to arise solely from multijet
production. For each sideband, a transfer factor per (\njet, \scalht)
bin is obtained from simulation, defined as the ratio of the number of
multijet events that satisfies the SR requirements to the number that
satisfies the sideband requirement. Estimates of the multijet
background per (\njet, \scalht) bin are obtained per sideband from the
product of the transfer factors and the corrected data counts.

The final estimate per (\njet, \scalht) bin is a weighted sum of the
three estimates. The multijet background is found to be small,
typically at the percent level, relative to the sum of all nonmultijet
backgrounds in all (\njet, \nb) bins of the SR. The \mhtmet and \bdphi
variables that are used to define the sidebands are determined to be
only weakly correlated for multijet events, and the estimates from
each sideband are assumed to be uncorrelated. Statistical
uncertainties associated with the finite event counts in data and
simulated event samples, as large as ${\approx}100\%$, are propagated
to each estimate. Uncertainties as large as ${\approx}20\%$ in the
estimates of nonmultijet contamination are also propagated to the
corrected events. Any differences between the three estimates per
(\njet, \scalht) bin are adequately covered by systematic
uncertainties of 100\%, which are assumed to be uncorrelated across
(\njet, \scalht) bins.

A model is assumed to determine the estimates as a function of \nb and
\mht. The distribution of multijet events as a function of \nb and
\mht per (\njet, \scalht) bin is assumed to be identical to the
distribution expected for the nonmultijet backgrounds. This assumption
is based on simulation-based studies and is a valid simplification
given the magnitude of the multijet background relative to the sum of
all other SM backgrounds, as well as the magnitude of the statistical
and systematic uncertainties in the estimates described above.

\section{Results}
\label{sec:result}

A likelihood model is used to obtain the SM expectations in the SR and
each CR, as well as to test for the presence of new-physics
signals. The observed event count in each bin, defined in terms of the
\njet, \nb, \scalht, and \mht variables, is modelled as a
Poisson-distributed variable around the SM expectation and a potential
signal contribution (assumed to be zero in the following
discussion). The expected event counts from nonmultijet processes in
the SR are related to those in the \mj and \mmj CRs via
simulation-based transfer factors, as described in
Section~\ref{sec:ewk}. The systematic uncertainties in the nonmultijet
estimates, summarized in Table~\ref{tab:bkgd_systs}, are accommodated
in the likelihood model as nuisance parameters, the measurements of
which are assumed to follow a log-normal distribution. In the case of
the modelling of the \mht distribution, alternative templates are used
to describe the uncertainties in the modelling and a vertical template
morphing schema~\cite{Prosper:2011zz, Khachatryan:2016dvc} is used to
interpolate between the nominal and alternative templates. The
multijet background estimates, determined using the method described
in Section~\ref{sec:qcd}, are also included in the likelihood model.

\begin{figure}[!p]
  \centering
  \includegraphics[width=0.99\textwidth]{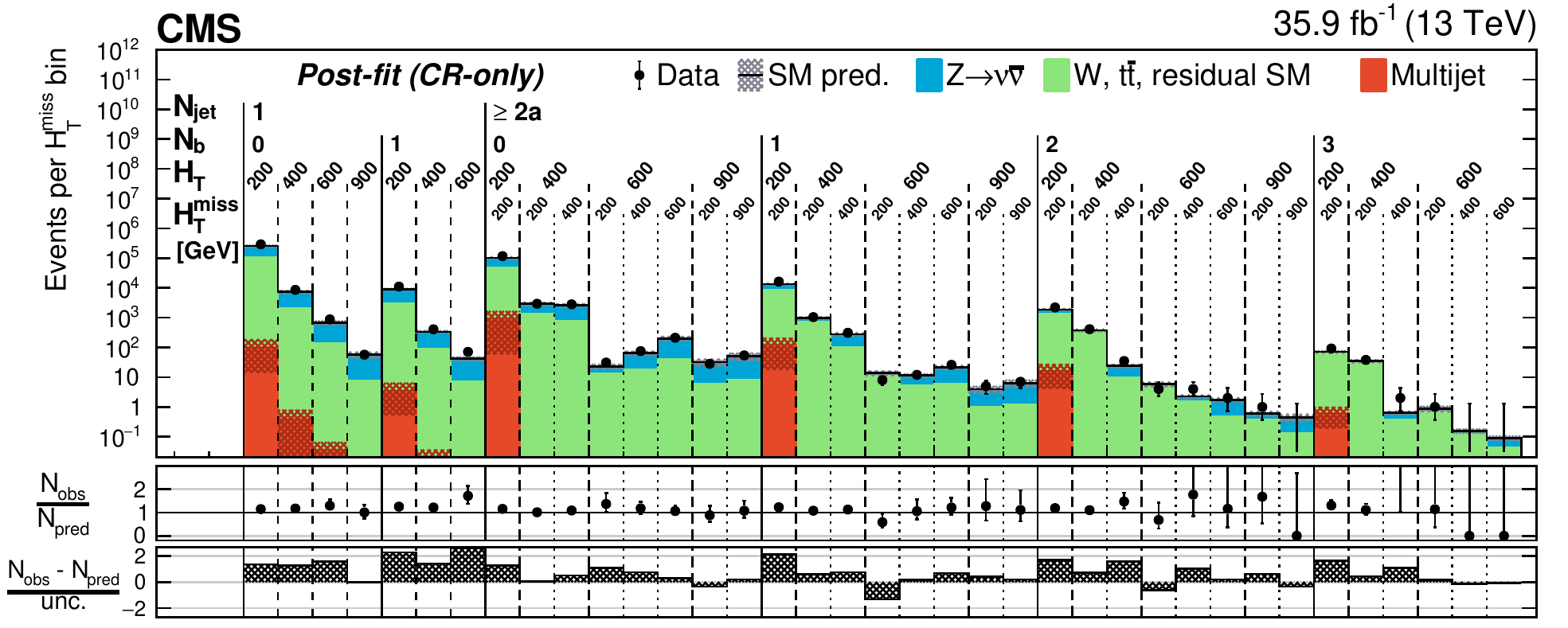}\\
  \includegraphics[width=0.99\textwidth]{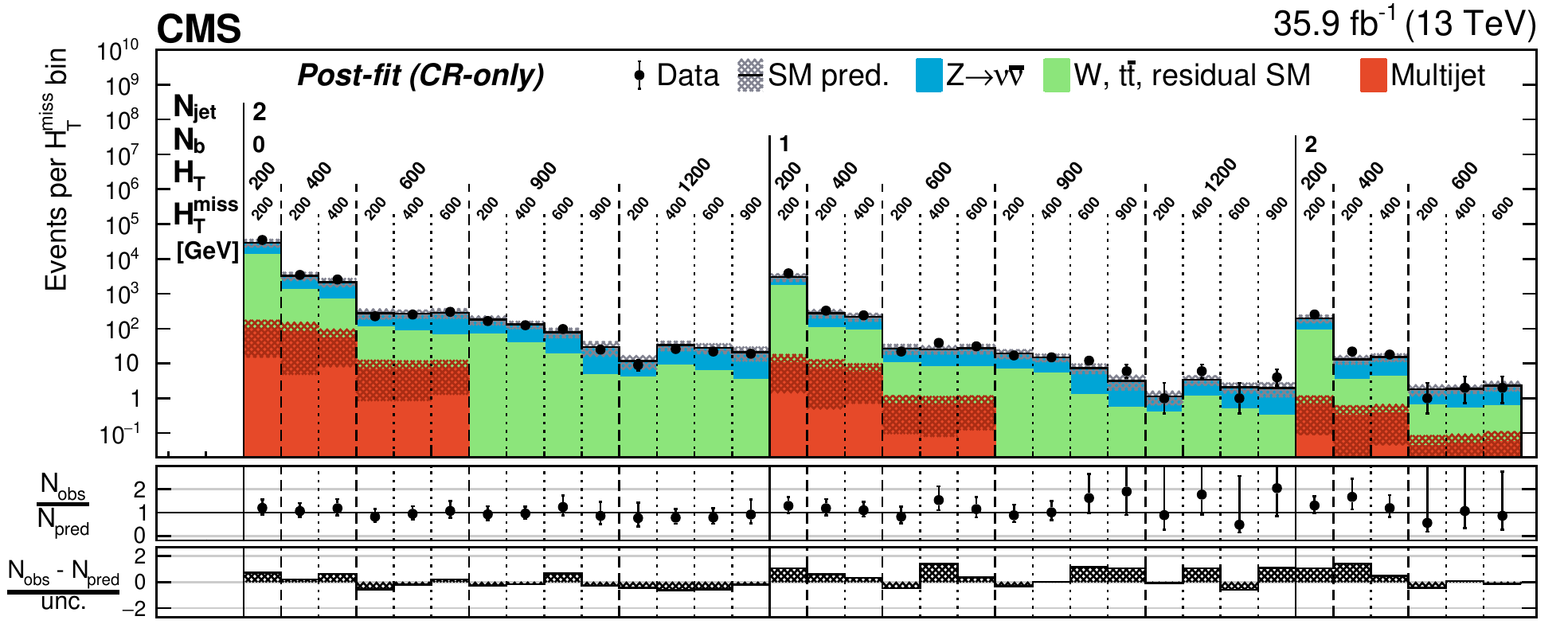}\\
  \includegraphics[width=0.99\textwidth]{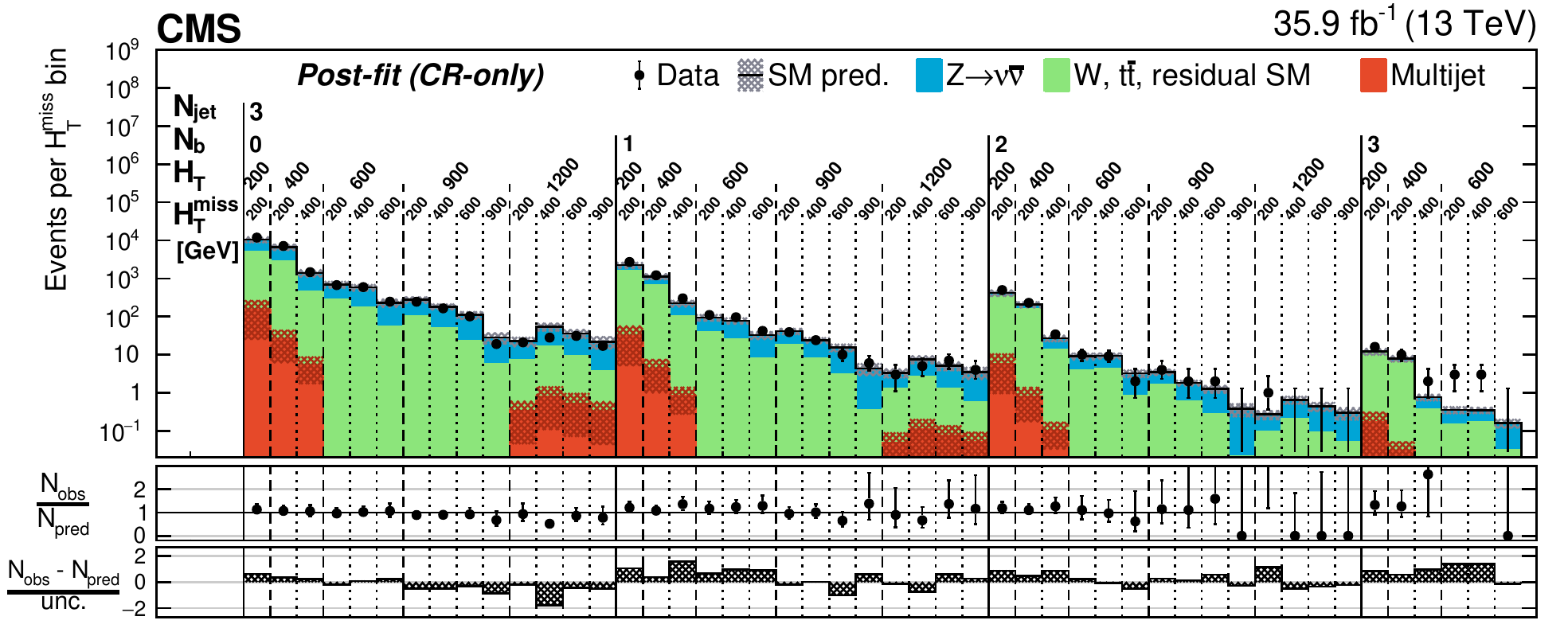}\\
  \caption{Counts of signal events (solid markers) and SM expectations
    with associated uncertainties (statistical and systematic, black
    histograms and shaded bands) as determined from the CR-only fit as
    a function of \nb, \scalht, and \mht for the event categories
    $\njet = 1$ and ${\geq}2a$ (upper), $=2$ (middle), and $=3$
    (lower). The centre panel of each subfigure shows the ratios of
    observed counts and the SM expectations, while the lower panel
    shows the significance of deviations observed in data with respect
    to the SM expectations expressed in terms of the total uncertainty
    in the SM expectations.  }
  \label{fig:result1}
\end{figure}

\begin{figure}[!p]
  \centering
  \includegraphics[width=0.99\textwidth]{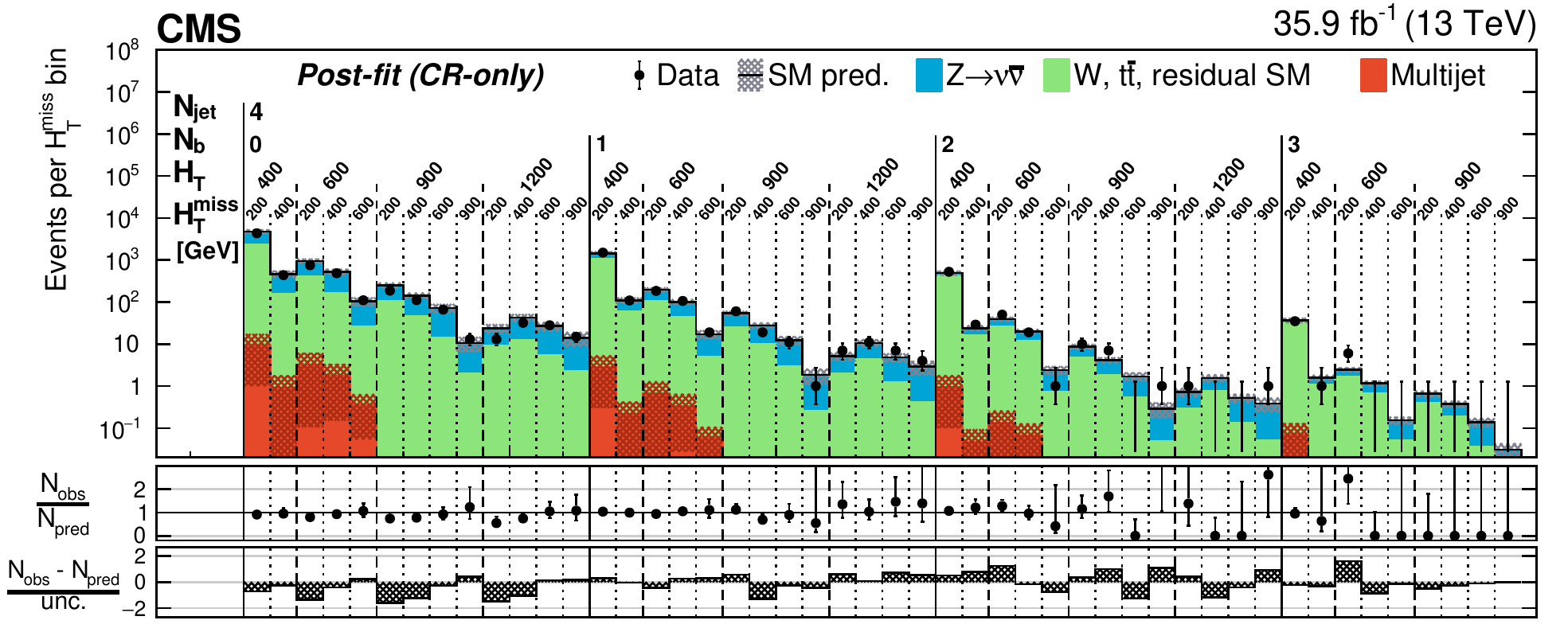}\\
  \includegraphics[width=0.99\textwidth]{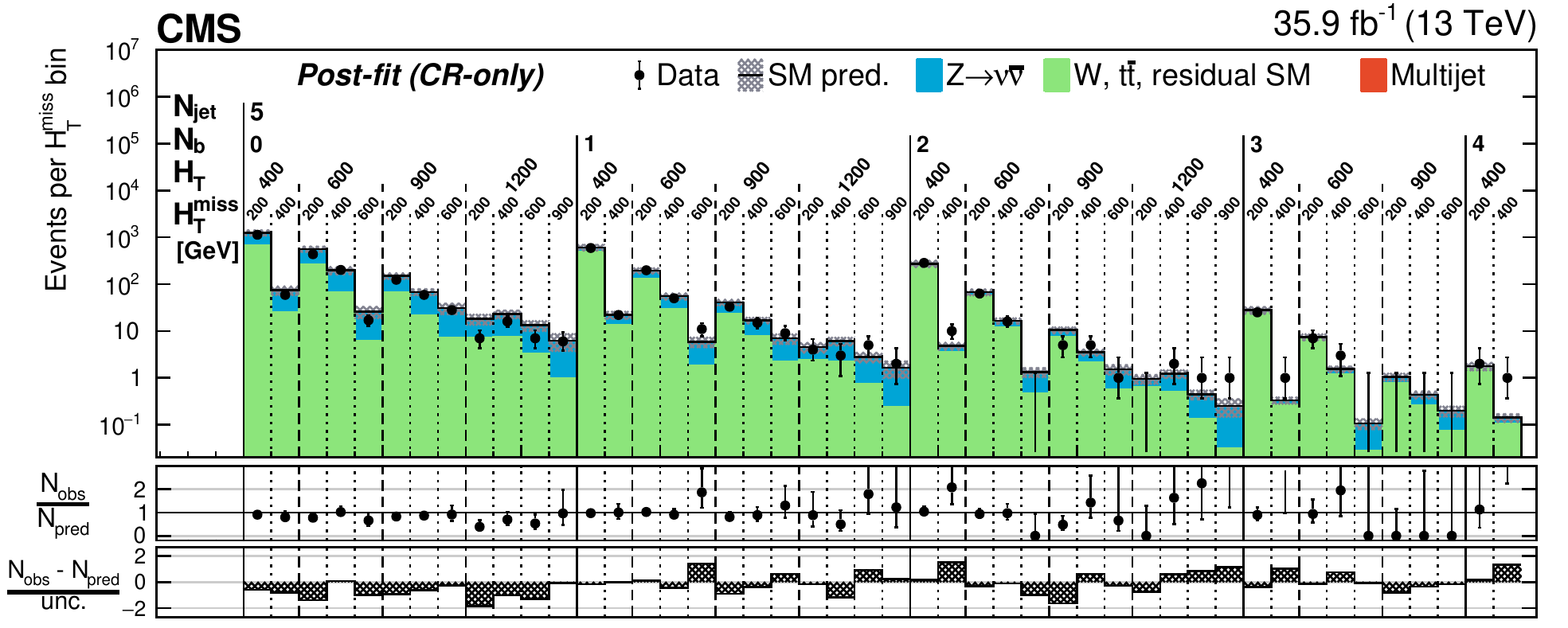}\\
  \includegraphics[width=0.99\textwidth]{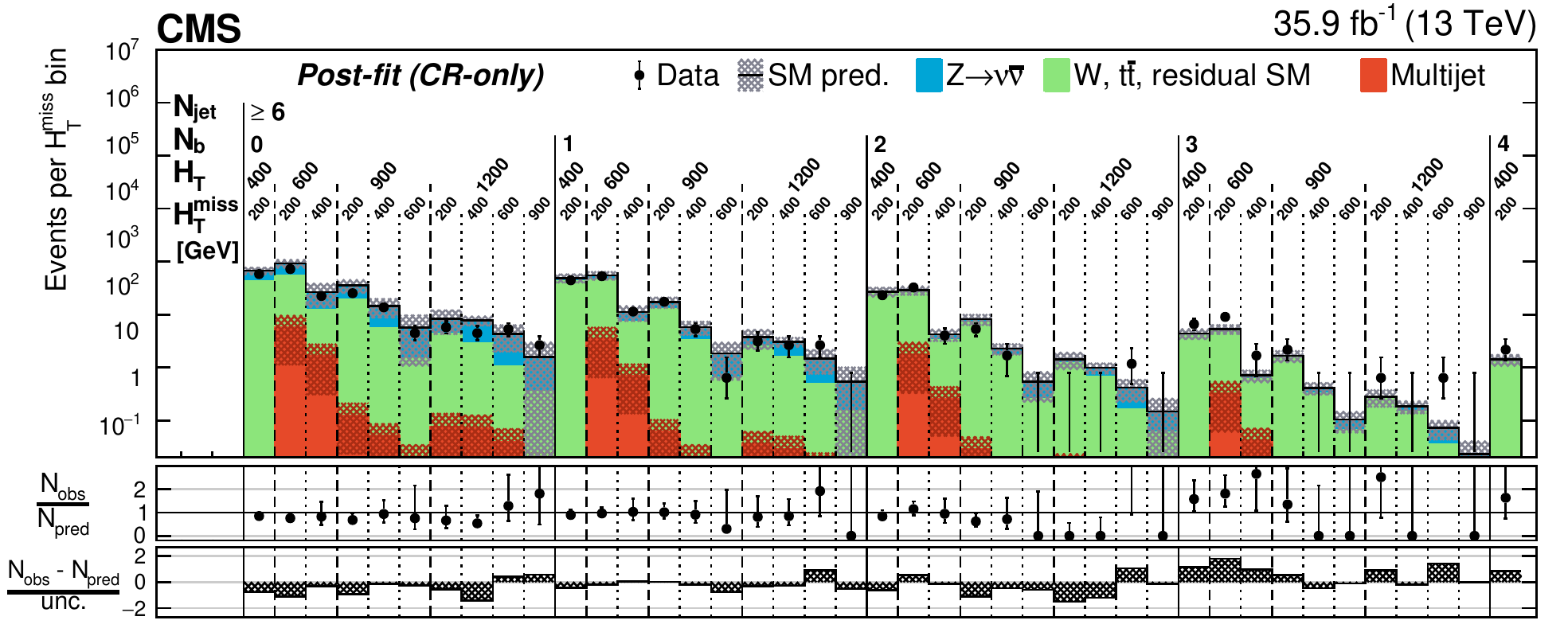}\\
  \caption{Counts of signal events (solid markers) and SM expectations
    with associated uncertainties (statistical and systematic, black
    histograms and shaded bands) as determined from the CR-only fit as
    a function of \nb, \scalht, and \mht for the event categories
    $\njet = 4$ (upper), $=5$ (middle), and ${\geq}6$ (lower). The
    lower panels are described in the caption of
    Fig.~\ref{fig:result1}.  }
  \label{fig:result2}
\end{figure}

Figures~\ref{fig:result1} and \ref{fig:result2} summarize the binned
counts of signal events and the corresponding SM expectations as
determined from a ``CR-only'' fit that uses only the data counts in
the \mj and \mmj control regions to constrain the model parameters
related to the nonmultijet backgrounds. The uncertainties in the SM
expectations reflect both statistical and systematic components. The
multijet background estimates are determined independently and
included in the SM expectations. The fit does not consider the event
counts in the signal region.

Hypothesis testing with regards to a potential signal contribution is
performed by considering a full fit to the event counts in the SR and
CRs. No significant deviation is observed between the predictions and
data in the SR and CRs, and the data counts appear to be adequately
modelled by the SM expectations with no significant kinematical
patterns.

Event counts, SM background estimates, and the associated correlation
matrix are also determined using the simplified 32-bin schema, which
can be found in Appendix~\ref{app:suppMat}.

\section{Interpretations}
\label{sec:interpretations}

The search result is used to constrain the parameter spaces of
simplified SUSY models~\cite{Alwall:2008ag, Alwall:2008va,
  sms}. Interpretations are provided for nine unique model families,
as summarized in Table~\ref{tab:sms}. Each family of models realizes a
unique production and decay mode. The model parameters are the masses
of the parent gluino ($m_\PSg$) or bottom, top, and LF ($m_{\PSQb_1}$,
$m_{\PSQt_1}$, $m_{\PSQ}$) squark, also collectively labelled as
$m_\text{SUSY}$, and the \PSGczDo ($m_{\PSGczDo}$). Two scenarios are
considered for LF squarks: one with an eightfold mass degeneracy for
$\PSQ_\cmsSymbolFace{L}$ and $\PSQ_\cmsSymbolFace{R}$ with $\PSQ =
\{\PSQu, \PSQd, \PSQs, \PSQc\}$ and the other with just a single light
squark ($\PSQu_\cmsSymbolFace{L}$). All other SUSY particles are
assumed to be too heavy to be produced directly. Gluinos are assumed
to undergo prompt three-body decays via highly virtual squarks. In the
case of split SUSY models (T1qqqqLL), the gluino is assumed to be
long-lived with proper decay lengths in the range $10^{-3} < \ctau <
10^{5}\unit{mm}$. A scenario involving a metastable gluino with $\ctau
= 10^{18}\unit{mm}$ is also considered.

Under the signal+background hypothesis, and in the presence of a
nonzero signal contribution, a modified frequentist approach is used
to determine observed upper limits (ULs) at 95\% confidence level (\CL)
on the cross section $\sigma_\text{UL}$ to produce pairs of SUSY
particles as a function of $m_\text{SUSY}$, $m_{\PSGczDo}$, and \ctau
(if applicable). The approach is based on the profile likelihood ratio
as the test statistic~\cite{CMS-NOTE-2011-005}, the \cls
criterion~\cite{junk, read}, and the asymptotic
formulae~\cite{Cowan:2010js} to approximate the distributions of the
test statistic under the SM-background-only and signal+background
hypotheses.  An Asimov data set~\cite{Cowan:2010js} is used to
determine the expected $\sigma_\text{UL}$ on the allowed cross section
for a given model. Potential signal contributions to event counts in
all bins of the SR and CRs are considered.

\begingroup
\renewcommand*{\arraystretch}{1.2}
\begin{table}[!t]
  \topcaption{Summary of the simplified SUSY models used to
    interpret the result of this search.}
  \label{tab:sms}
  \centering
  \begin{tabular}{ lll }
    \hline
    Model family
    & Production and decay
    & Additional assumptions                                                         \\
    \hline
    \multicolumn{3}{l}{\textbf{Production and prompt decay of squark pairs} }          \\
    T2bb
    & $\Pp\Pp \to \PSQb_1\PASQb_1$,
    $\PSQb_1 \to \cPqb\PSGczDo$
    & \NA                                                                            \\
    T2tt
    & $\Pp\Pp \to \PSQt_1\PASQt_1$,
    $\PSQt_1 \to \cPqt\PSGczDo$
    & \NA                                                                            \\
    T2cc
    & $\Pp\Pp \to \PSQt_1\PASQt_1$,
    $\PSQt_1\to \cPqc\PSGczDo$
    & $10 < m_{\,\PSQt_1} - m_{\PSGczDo} < 80\GeV$                                   \\
    T2qq\_8fold
    & $\Pp\Pp \to \PSQ\PASQ$,
    $\PSQ \to \cPq\PSGczDo$
    & $m_{\PSQ_\cmsSymbolFace{L}} = m_{\PSQ_\cmsSymbolFace{R}}$,
    $\PSQ = \{ \PSQu, \PSQd, \PSQs, \PSQc \}$                                     \\
    T2qq\_1fold
    & $\Pp\Pp \to \PSQ\PASQ$,
    $\PSQ \to \cPq\PSGczDo$
    & $m_{\PSQ (\PSQ \neq \PSQu_\cmsSymbolFace{L})} \gg m_{\PSQu_\cmsSymbolFace{L}}$ \\
    \multicolumn{3}{l}{\textbf{Production and prompt decay of gluino pairs} }          \\
    T1bbbb
    & $\Pp\Pp \to \PSg\PSg$,
    $\PSg\to \cPaqb\PSQb_1^* \to \cPaqb\cPqb\PSGczDo$
    & $m_{\PSQb_1} \gg m_{\PSg}$                                                     \\
    T1tttt
    & $\Pp\Pp \to \PSg\PSg$,
    $\PSg\to \cPaqt\PSQt_1^* \to \cPaqt\cPqt\PSGczDo$
    & $m_{\PSQt_1} \gg m_{\PSg}$                                                     \\
    T1qqqq
    & $\Pp\Pp \to \PSg\PSg$,
    $\PSg\to \cPaq\PSQ^* \to \cPaq\cPq\PSGczDo$
    & $m_{\PSQ} \gg m_{\PSg}$                                                        \\
    \multicolumn{3}{l}{\textbf{Production and decay of long-lived gluino pairs}}       \\
    T1qqqqLL
    & $\Pp\Pp \to \PSg\PSg$,
    $\PSg \to \cPaq\PSQ^* \to \cPaq\cPq\PSGczDo$
    & $m_{\PSQ} \gg m_{\PSg}$, $10^{-3} < \ctau < 10^{5}\unit{mm}$ or metastable     \\
    \hline
  \end{tabular}
\end{table}
\endgroup

The experimental acceptance times efficiency (\ate) is evaluated
independently for each model, defined in terms of $m_\text{SUSY}$,
$m_{\PSGczDo}$, and \ctau (if applicable). The effects of several
sources of uncertainty in \ate, as well as the potential for migration
of events between bins of the SR, are considered. Correlations are
taken into account where appropriate, including those relevant to
signal contamination that may contribute to counts in the CRs.

The statistical uncertainty arising from the finite size of simulated
samples can be as large as ${\approx}30\%$. The \ate for models with a
compressed mass spectrum relies on jets arising from ISR, the
modelling of which is evaluated using the technique described in
Ref.~\cite{Chatrchyan:2013xna}. The associated uncertainty can be as
large as ${\approx}30\%$. The corrections to the jet energy scale
(JES) evaluated with simulated events can lead to variations in event
counts as large as ${\approx}25\%$ for models yielding high jet
multiplicities. The uncertainties in the modelling of scale factors
applied to simulated event samples that correct for differences in the
b tagging efficiency and mistag probabilities can be as large as
${\approx}20\%$.

Table~\ref{tab:benchmarks} defines a number of benchmark models that
are close to the limit of the search sensitivity. All model families
are represented, and the model parameters ($m_\text{SUSY}$,
$m_{\PSGczDo}$, and \ctau if applicable) are chosen to select models
with large and small differences in $m_\text{SUSY}$ and
$m_{\PSGczDo}$, as well as a range of \ctau
values. Table~\ref{tab:benchmarks} summarizes the aforementioned
uncertainties for each benchmark model, presented in terms of a
characteristic range that is representative of the variations observed
across the bins of the SR. The upper bound for each range may be
subject to moderate statistical fluctuations.

\begingroup
\renewcommand*{\arraystretch}{1.1}
\begin{table}[!t]
  \topcaption{A list of benchmark simplified models organized
    according to production and decay modes (family), the \ate,
    representative values for some of the dominant sources of
    systematic uncertainty, and the expected and observed upper limits
    on the production cross section $\sigma_\text{UL}$ relative to the
    theoretical value $\sigma_\text{th}$ calculated at NLO+NLL
    accuracy. Additional uncertainties concerning the T1qqqqLL models
    are not listed here and are discussed in the text.
  }
  \label{tab:benchmarks}
  \centering
  \resizebox{\textwidth}{!}{
    \begin{tabular}{ lrcrrrrrcc }
      \hline
      Family
      & $(m_{\text{SUSY}}, m_{\PSGczDo})$
      & \ate
      & \multicolumn{4}{c}{Systematic uncertainties [\%]}
      & \multicolumn{2}{c}{$\sigma_\text{UL}/\sigma_\text{th}$ (95\% \CL)} \\ [0.3ex]
      \cline{4-7}
      (\ctau)
      & [\GeVns{}]
      & [\%]
      & MC stat.
      & ISR
      & JES
      & b tagging
      & Exp.
      & Obs.                                                              \\ [0.3ex]
      \hline
      \multirow{2}{*}{T2bb}
      & (1000, 100)
      & 40.1           & 14--23 & 1--7   & 4--11  & 1--4
      & 0.62           & 0.67                                             \\
      & (550, 450)
      & \phantom{1}5.7 & 9--22  & 4--15  & 4--15  & 3--7
      & 0.76           & 1.21                                             \\ [0.5ex]
      \multirow{3}{*}{T2tt}
      & (1000, 50)
      & 23.8           & 14--27 & 3--7   & 4--14  & 1--5
      & 0.82           & 0.85                                             \\
      & (450, 200)
      & \phantom{1}4.2 & 6--19  & 4--12  & 6--15  & 4--9
      & 0.56           & 0.73                                             \\ [0.5ex]
      & (250, 150)
      & \phantom{1}0.3 & 10--23 & 13--27 & 8--22  & 6--16
      & 0.71           & 0.66                                             \\ [0.5ex]
      \multirow{1}{*}{T2cc}
      & (500, 480)
      & 20.5           & 6--19  & 4--18  & 5--13  & 1--4
      & 0.68           & 1.38                                             \\ [0.5ex]
      \multirow{2}{*}{T2qq\_8fold}
      & (1250, 100)
      & 42.9           & 12--24 & 2--7   & 5--14  & 1--1
      & 0.54           & 0.66                                             \\
      & (700, 600)
      & \ph{1}7.7      & 6--22  & 4--17  & 4--13  & 2--5
      & 0.75           & 1.13                                             \\ [0.5ex]
      \multirow{2}{*}{T2qq\_1fold}
      & (700, 100)
      & 32.9           & 4--22  & 2--7   & 3--10  & 0--5
      & 0.60           & 0.88                                             \\
      & (400, 300)
      & \ph{1}4.5      & 6--20  & 5--22  & 5--18  & 3--5
      & 0.61           & 0.46                                             \\ [0.5ex]
      \multirow{2}{*}{T1bbbb}
      & (1900, 100)
      & 25.1           & 11--19 & 3--9   & 4--6   & 7--11
      & 0.56           & 1.25                                             \\
      & (1300, 1100)
      & 14.6           & 11--22 & 2--11  & 3--11  & 2--5
      & 0.44           & 1.15                                             \\ [0.5ex]
      \multirow{2}{*}{T1tttt}
      & (1700, 100)
      & \phantom{1}6.9 & 12--24 & 2--6   & 3--15  & 2--6
      & 0.51           & 1.31                                             \\
      & (950, 600)
      & \phantom{1}0.3 & 15--30 & 5--9   & 12--26 & 2--6
      & 0.89           & 1.51                                             \\ [0.5ex]
      T1qqqqLL
      & (1800, 200)
      & 27.8           & 8--20  & 3--5   & 3--9   & 0--1
      & 1.02           & 1.91                                             \\
      ($1\um$)
      & (1000, 900)
      & \ph{1}6.7      & 15--21 & 2--10  & 4--14  & 0--1
      & 0.68           & 1.26                                             \\ [0.5ex]
      T1qqqqLL
      & (1800, 200)
      & 22.9           & 11--20 & 2--5   & 3--9   & 17--59
      & 0.43           & 1.00                                             \\
      ($1\unit{mm}$)
      & (1000, 900)
      & \ph{1}5.2      & 17--26 & 2--9   & 4--17  & 10--41
      & 0.28           & 0.63                                             \\ [0.5ex]
      T1qqqqLL
      & (1000, 200)
      & 11.2           & 16--22 & 2--14  & 4--9   & 0--1
      & 0.74           & 1.58                                             \\
      ($100\unit{m}$)
      & (1000, 900)
      & 10.4           & 14--26 & 3--14  & 2--12  & 0--1
      & 0.63           & 0.45                                             \\ [0.5ex]
      \hline
    \end{tabular}
  }
\end{table}
\endgroup

Additional subdominant contributions to the total uncertainty are also
considered. The uncertainty in the integrated luminosity is determined
to be 2.5\%~\cite{CMS:2017sdi}. Uncertainties in the production cross
section arising from the choice of the PDF set, and variations
therein, as well as variations in $\mu_\mathrm{F}$ and $\mu_\mathrm{R}$ at
LO are considered. Uncertainties in event migration between bins from
variations in the PDF sets are assumed to be correlated with, and
adequately covered by, the uncertainties in the modelling of
ISR. Uncertainties from $\mu_\mathrm{F}$ and $\mu_\mathrm{R}$ variations
are determined to be ${\approx}5\%$. The effect of a 5\% uncertainty
in the total inelastic cross section~\cite{Aaboud:2016mmw} is
propagated through the weighting procedure that corrects for
differences between the simulated and measured pileup, resulting in
event count variations of ${\approx}10\%$. Uncertainty in the
modelling of the efficiency to identify high-quality, isolated leptons
is ${\approx}5\%$ and is treated as anticorrelated between the SR and
\mj and \mmj CRs. The uncertainty in the trigger efficiency to record
signal events is ${<}10\%$.

The \ate for the T1qqqqLL family of models depends on \ctau in
addition to $m_\PSg$ and $m_{\PSGczDo}$. Scenarios involving a
compressed mass spectrum or gluinos with $\ctau \gtrsim 10\unit{m}$
increase the probability that the decay of the gluino-pair system
escapes detection, and the \ate is reduced for such models, as
indicated in Table~\ref{tab:benchmarks}, because of an increased
reliance on jets from ISR. Scenarios with $m_\PSg - m_{\PSGczDo}
\gtrsim 100\GeV$ and $1 \lesssim \ctau \lesssim 10\unit{m}$ often lead
to one or both gluinos decaying within the calorimeter systems to
yield energetic jets comprising particle candidates that have no
associated charged particle track. Hence, the efficiencies for the
event vetoes related to the jet identification and \fhleadjet
requirements, described in Sections~\ref{sec:reconstruction} and
\ref{sec:baseline}, can be as low as ${\approx}90\%$ and
${\approx}30\%$, respectively, for this region of the model parameter
space. Uncertainties as large as ${\approx}10\%$ are assumed. The
efficiencies for all other scenarios are typically
${\approx}100\%$. Jet identification requirements in the trigger logic
lead to inefficiencies and uncertainties not larger than 2\%. Finally,
models with $1 \lesssim \ctau \lesssim 10\unit{mm}$ often lead to jets
that are tagged by the CSV algorithm with efficiencies as high as
${\approx}60\%$, which are comparable to the values obtained for jets
originating from b quarks. Uncertainties of 20--50\% in the tagging
efficiency are assumed to cover differences with respect to jets
originating from b quarks, as indicated in Table~\ref{tab:benchmarks}.

\begin{figure}[!p]
  \centering
  \includegraphics[width=0.6\textwidth]{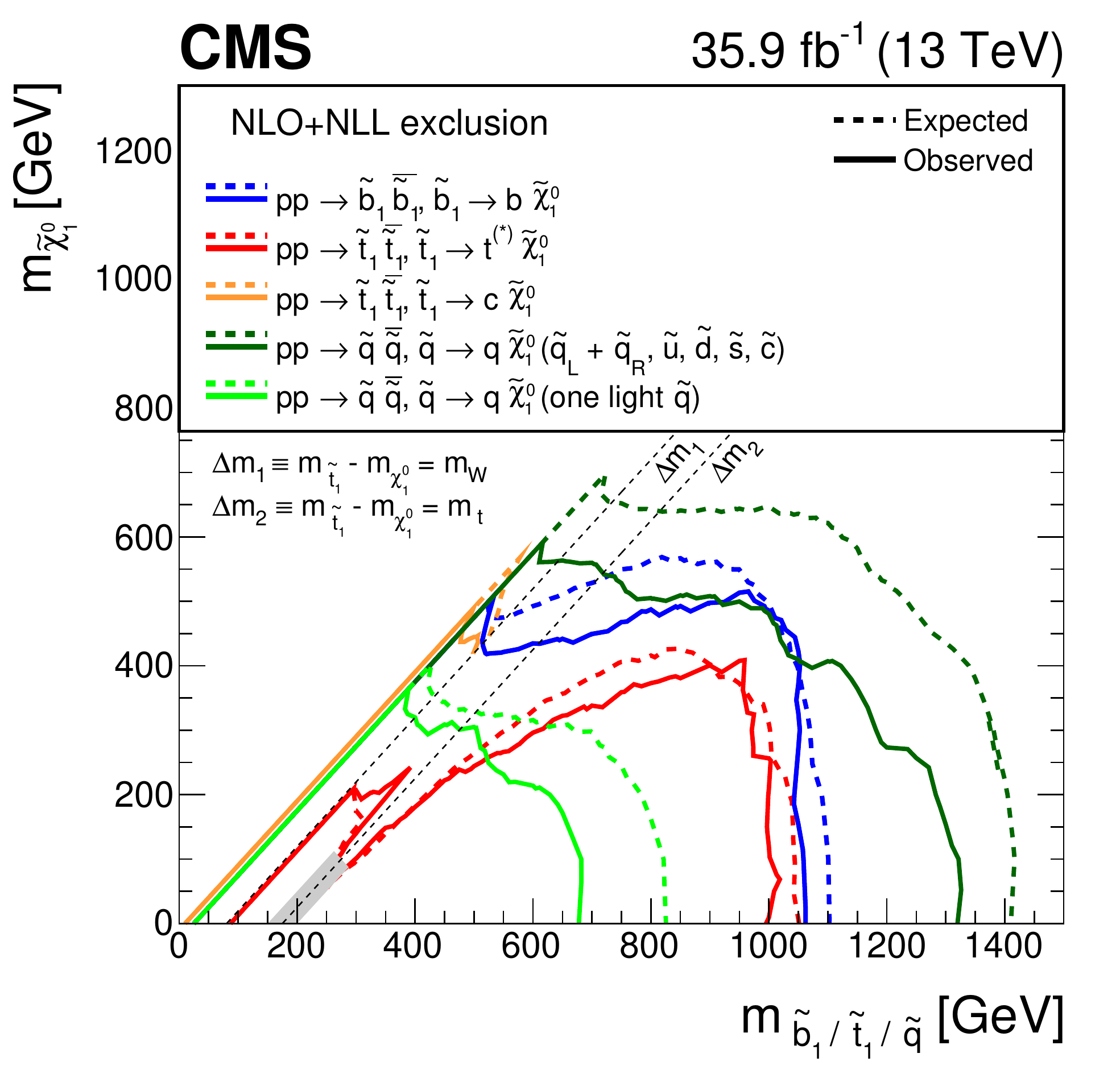}\\
  \includegraphics[width=0.6\textwidth]{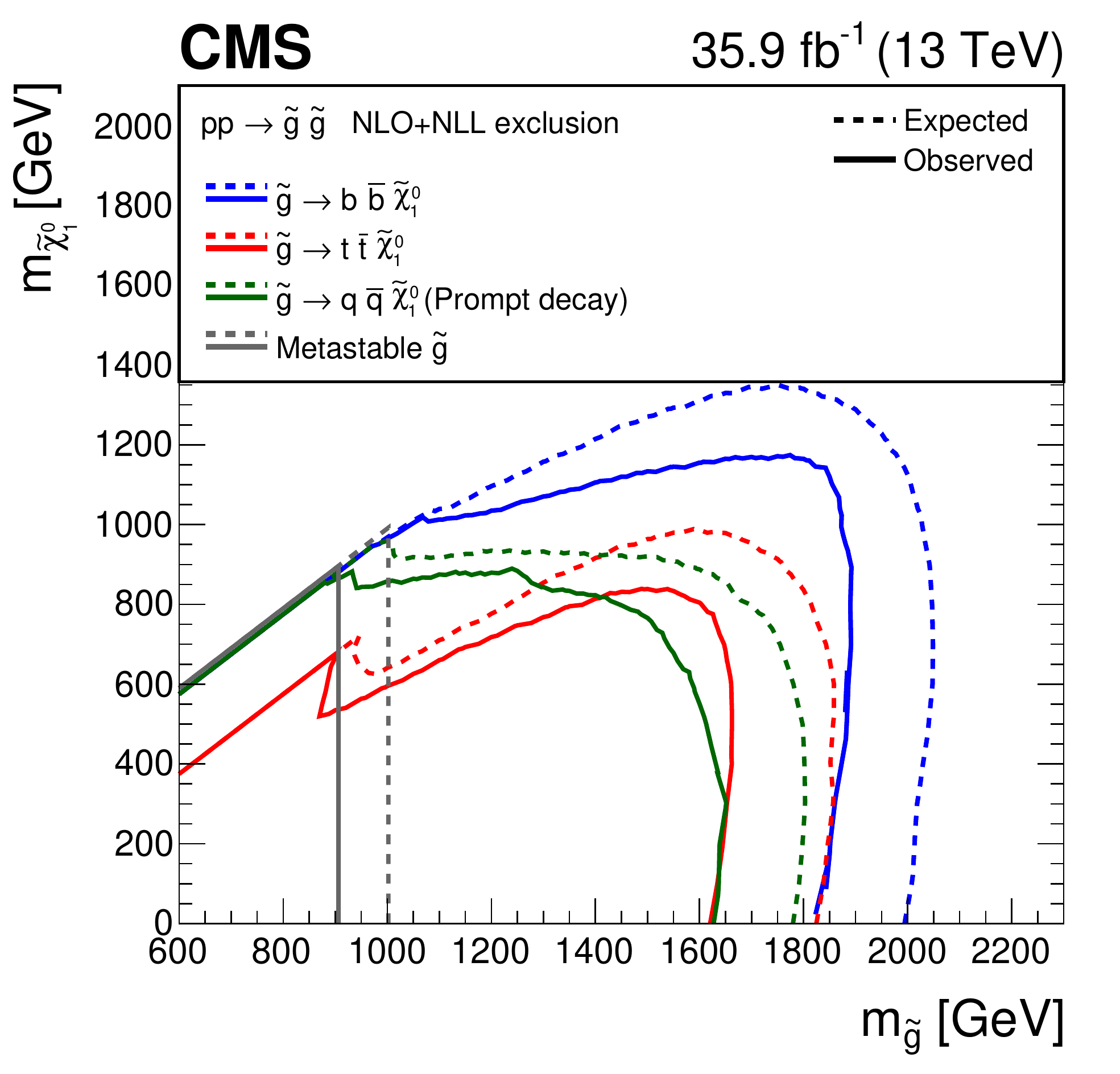}\\
  \caption{Observed and expected mass exclusions at 95\% \CL
    (indicated, respectively, by solid and dashed contours) for
    various families of simplified models. The upper subfigure
    summarises the mass exclusions for five model families that
    involve the direct pair production of squarks. The first scenario
    considers the pair production and decay of bottom squarks
    (T2bb). Two scenarios involve the production and decay of top
    squark pairs (T2tt and T2cc). The grey shaded region denotes T2tt
    models that are not considered for interpretation. Two further
    scenarios involve, respectively, the production and decay of
    light-flavour squarks, with different assumptions on the mass
    degeneracy of the squarks as described in the text (T2qq\_8fold
    and T2qq\_1fold).  The lower subfigure summarises three scenarios
    that involve the production and prompt decay of gluino pairs via
    virtual squarks (T1bbbb, T1tttt, and T1qqqq). A final scenario
    involves the production of gluinos that are assumed to be
    metastable on the detector scale (T1qqqqLL).}
  \label{fig:limits-sms}
\end{figure}

Figure~\ref{fig:limits-sms} summarizes the excluded regions of the
mass parameter space for the nine families of simplified models. The
regions are determined by comparing $\sigma_\text{UL}$ with the
theoretical cross sections $\sigma_\text{th}$ calculated at NLO+NLL
accuracy.  The former value is determined as a function of
$m_{\text{SUSY}}$ and $m_{\PSGczDo}$, while the latter has a
dependence solely on $m_{\text{SUSY}}$. The exclusion of models is
evaluated using observed data counts in the signal region (solid
contours) and also expected counts based on an Asimov data set (dashed
contours). The observed excluded regions for the T1bbbb and T1tttt
families, as shown in Fig.~\ref{fig:limits-sms} (lower), can be up to
2--3 standard deviations weaker than the expected excluded regions
when $m_{\PSg} - m_{\PSGczDo} \approx 350\GeV$. These differences are
typically due to fluctuations in data for events that satisfy $\njet
\geq 5$ and $\nb \geq 3$. Figure~\ref{fig:limits-sms} (lower) also
allows a comparison of the sensitivity to T1qqqq and T1qqqqLL models,
which assume the prompt-decay and metastable gluino scenarios,
respectively. The latter scenario leads to a monojet-like final state
as the gluino escapes detection, resulting in a reach in $m_\PSg$ that
is independent of $m_{\PSGczDo}$.

\begin{figure}[!p]
  \centering
  \includegraphics[width=0.33\textwidth]{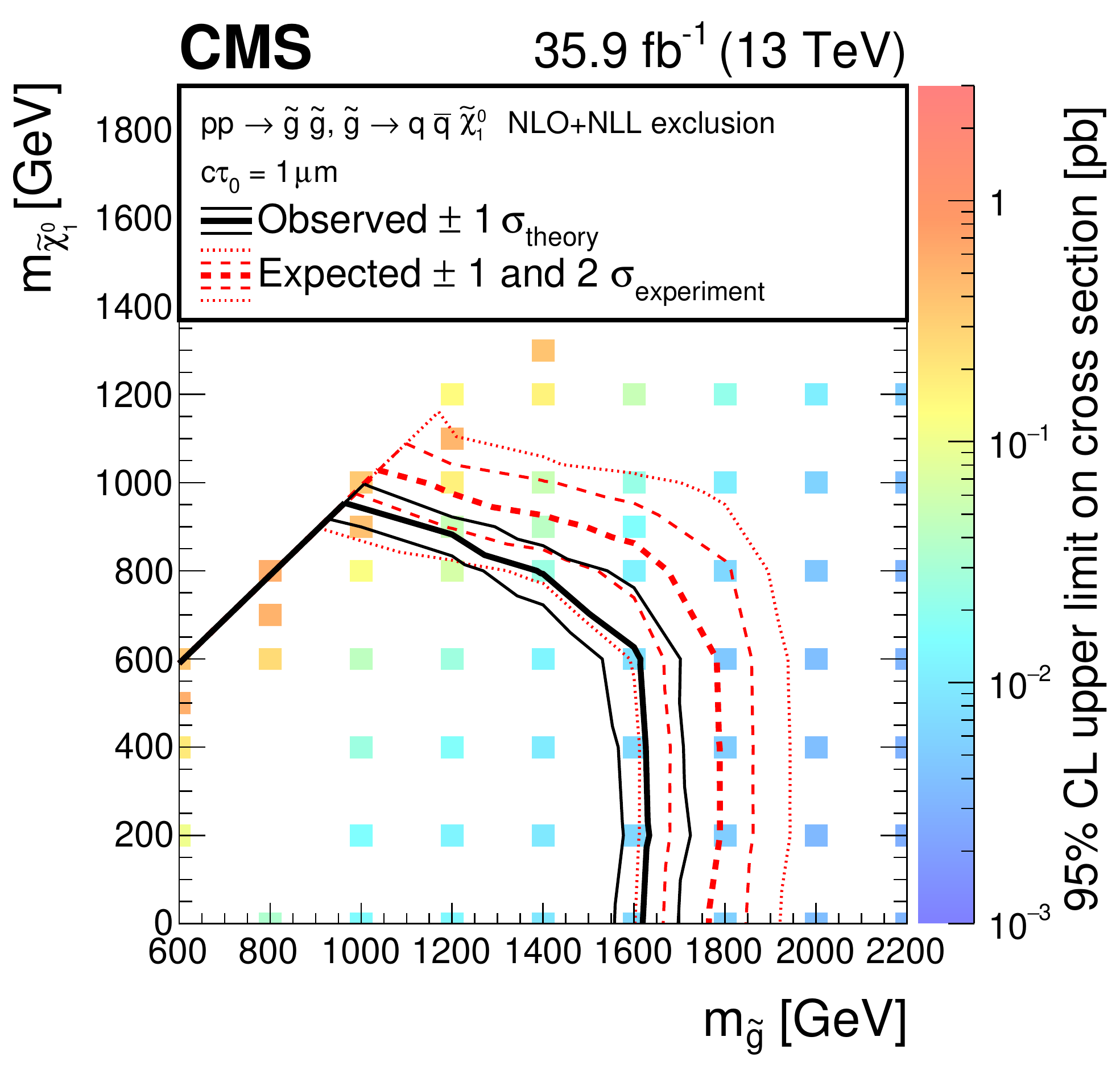}~
  \includegraphics[width=0.33\textwidth]{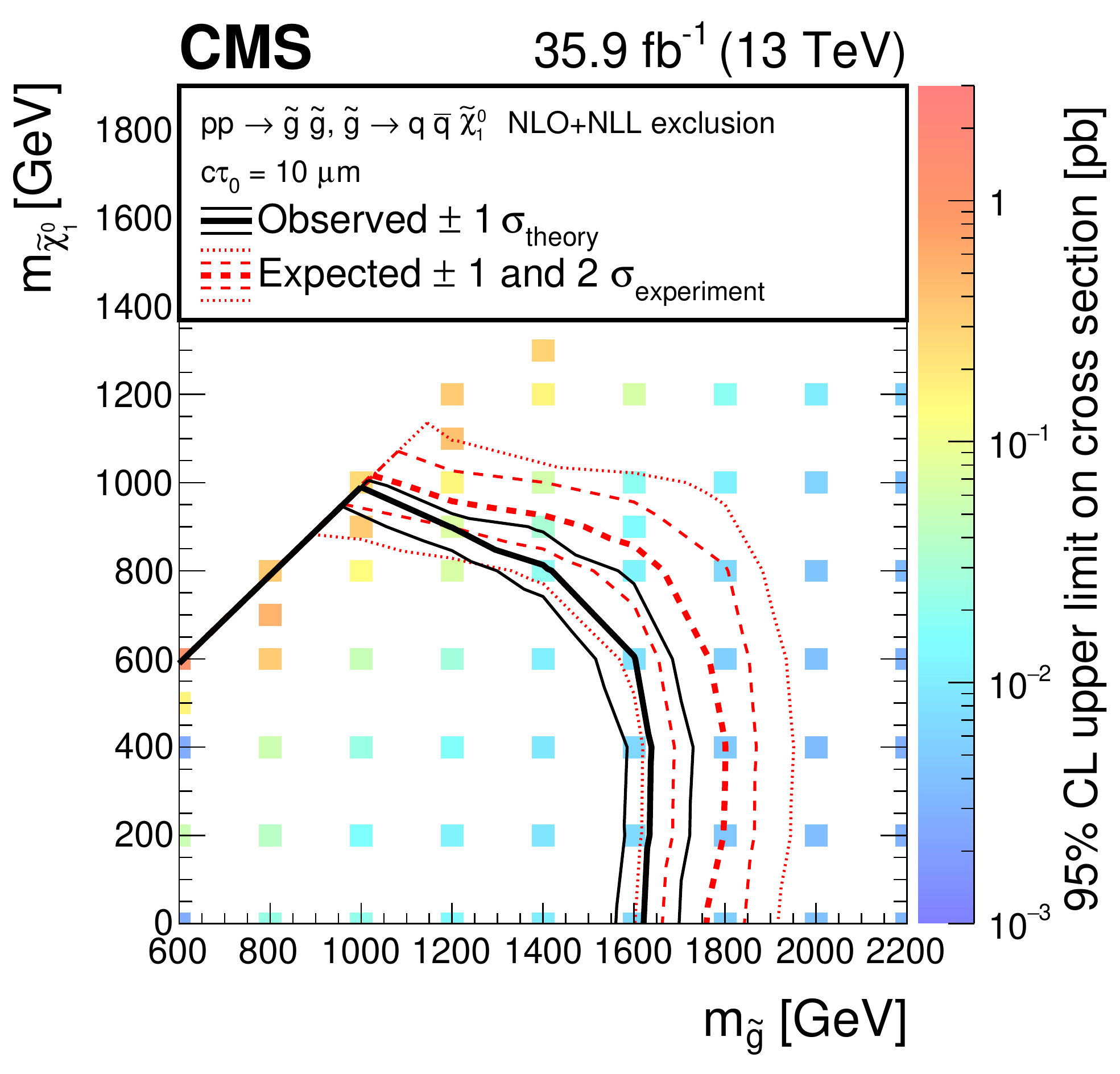}~
  \includegraphics[width=0.33\textwidth]{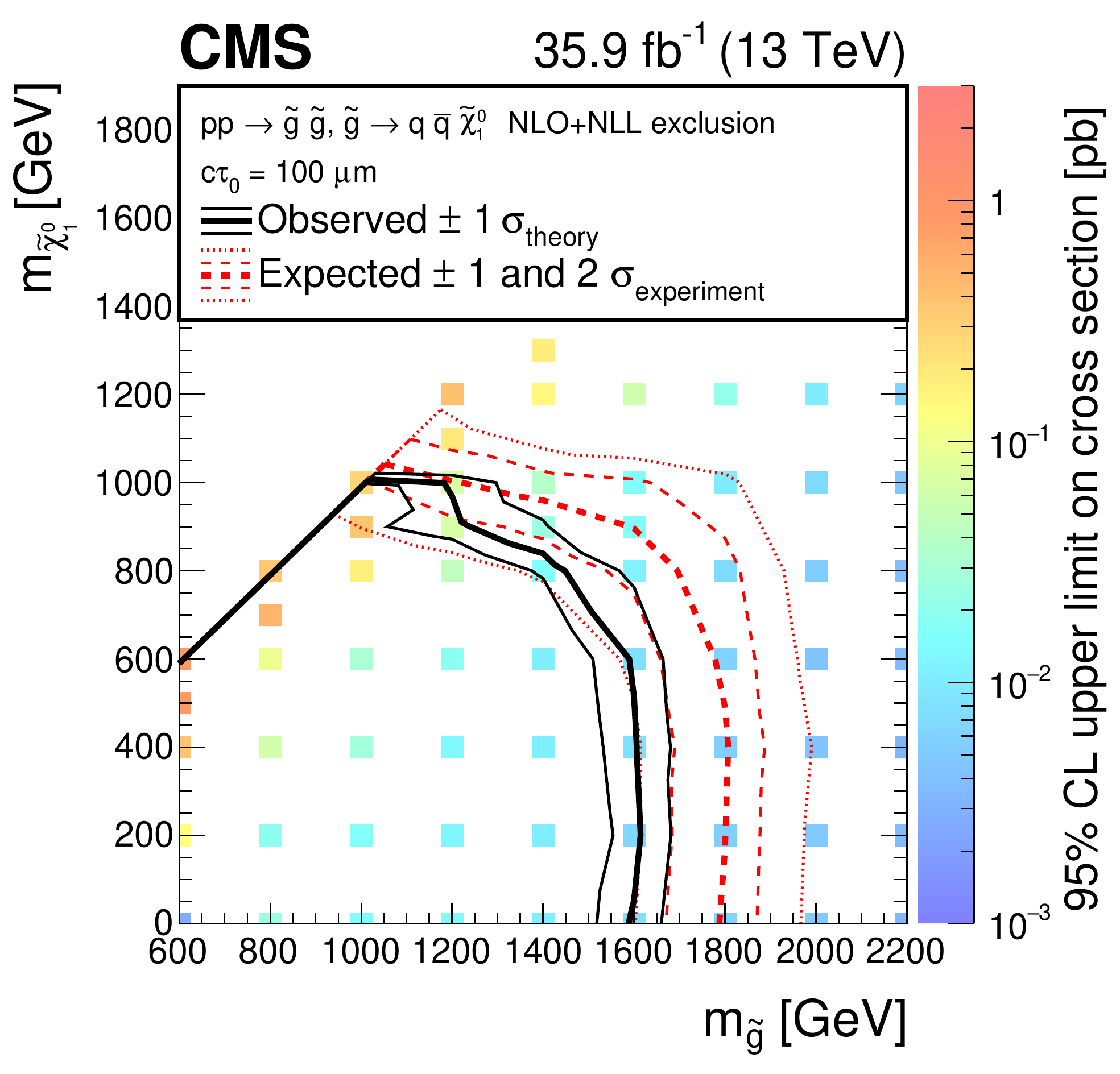}\\
  \includegraphics[width=0.33\textwidth]{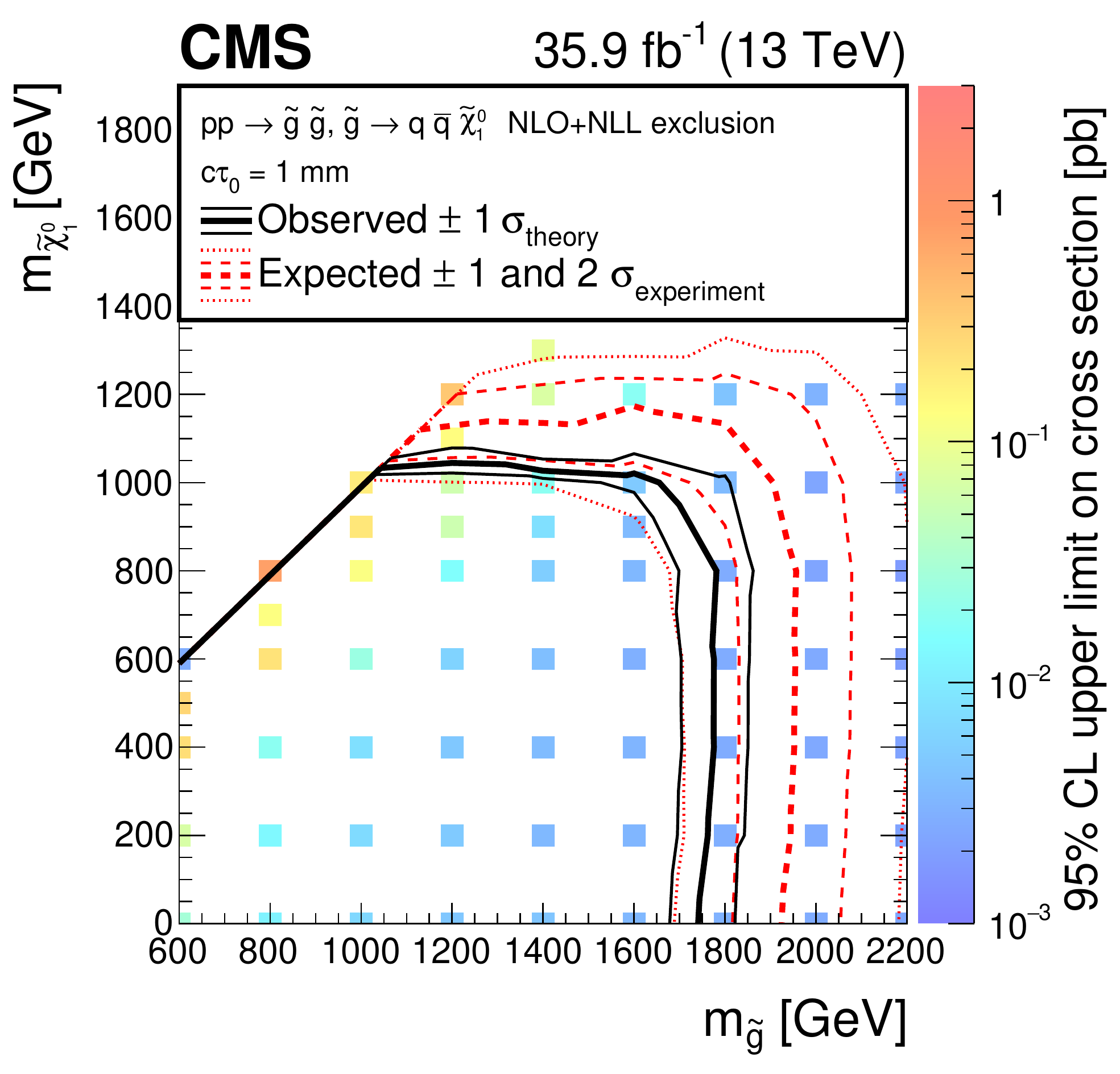}~
  \includegraphics[width=0.33\textwidth]{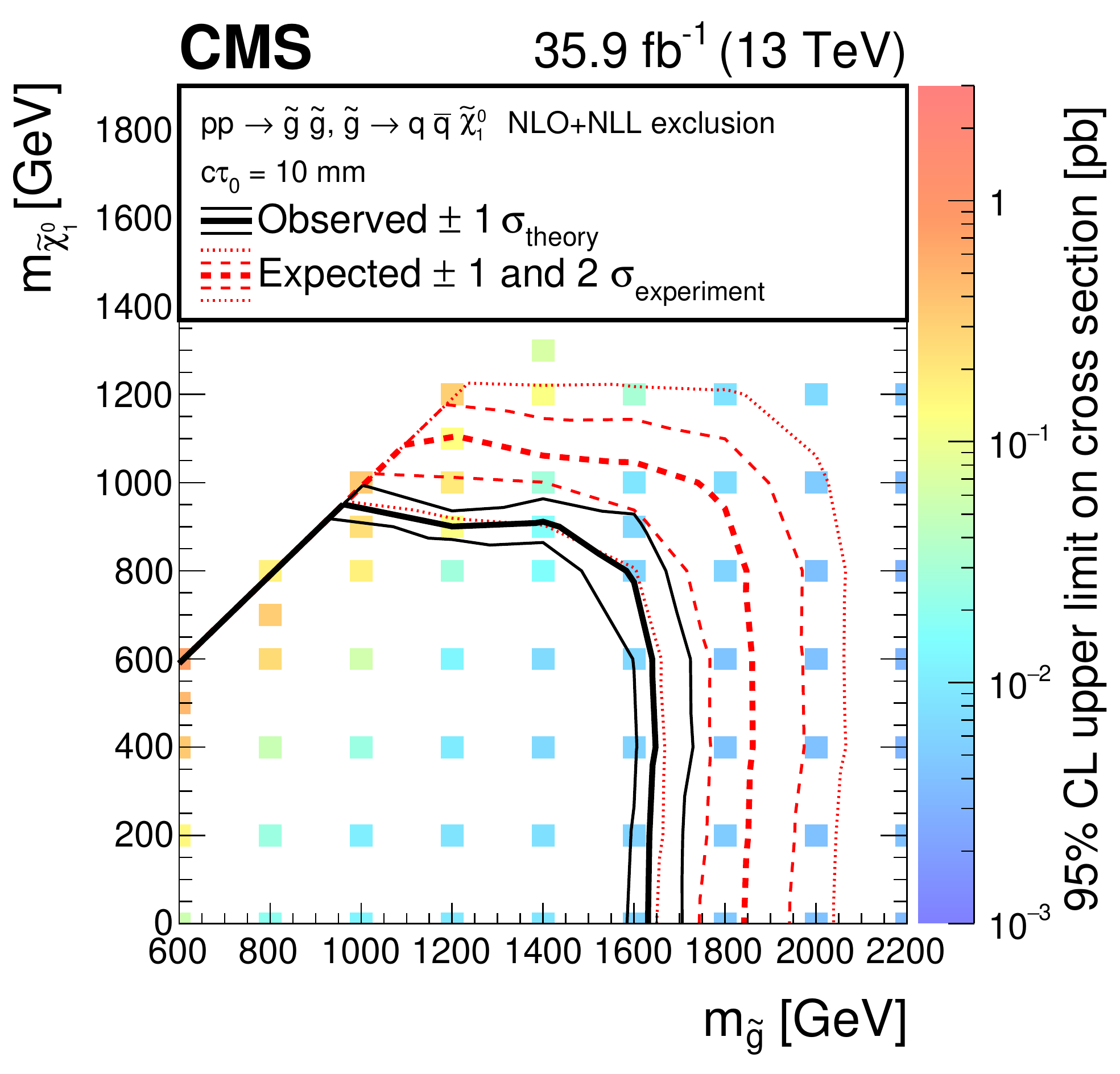}~
  \includegraphics[width=0.33\textwidth]{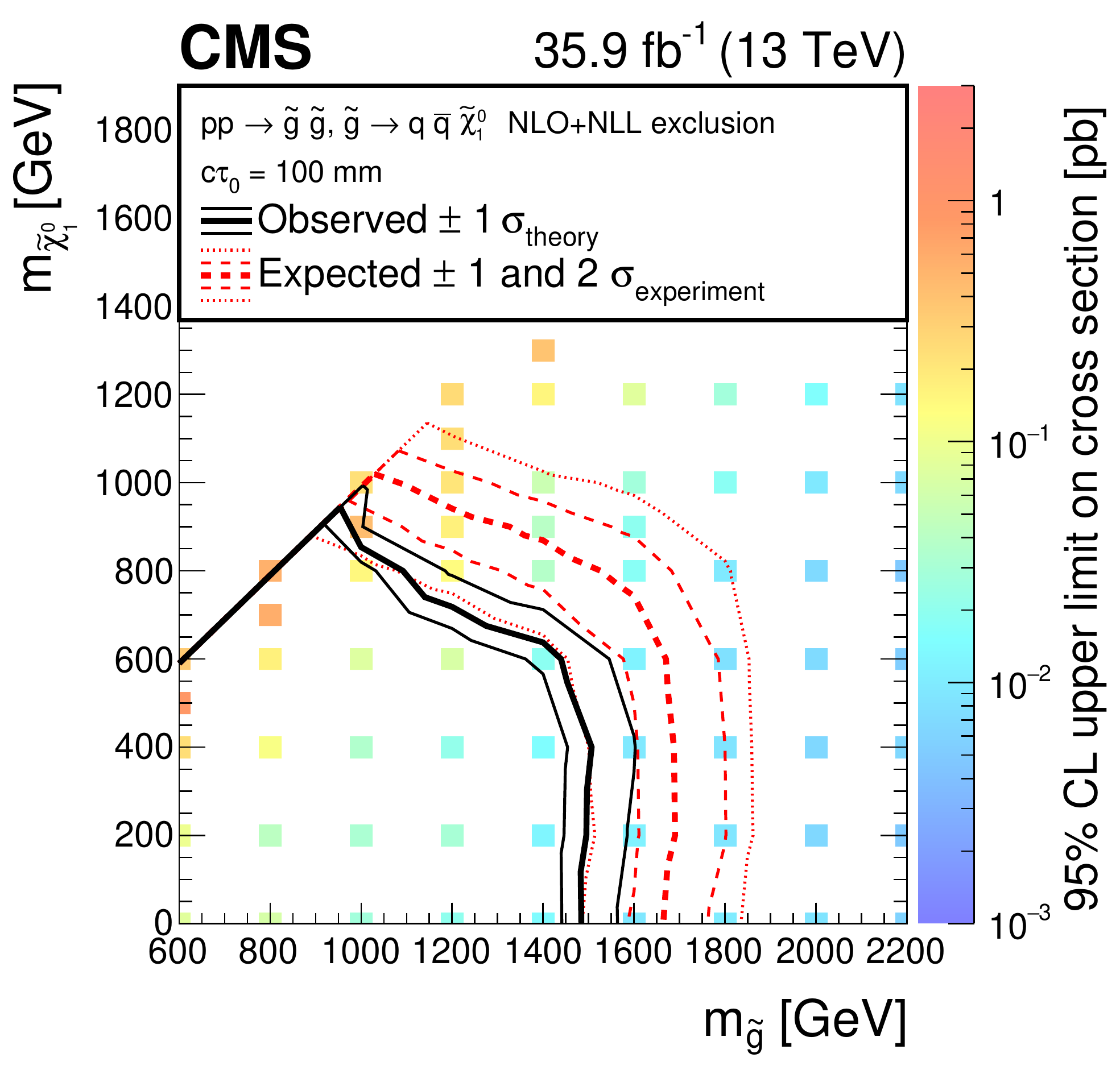}\\
  \includegraphics[width=0.33\textwidth]{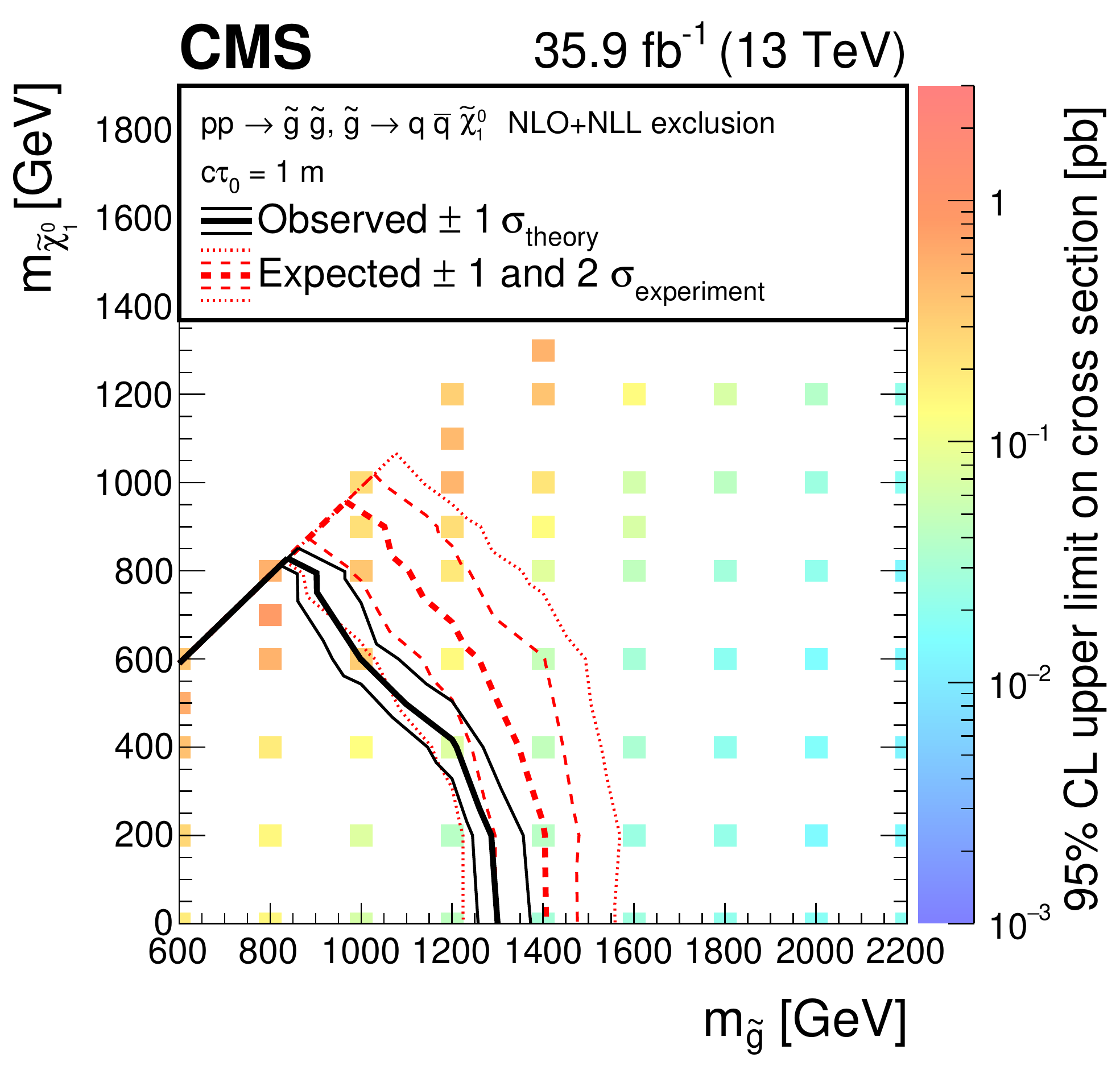}~
  \includegraphics[width=0.33\textwidth]{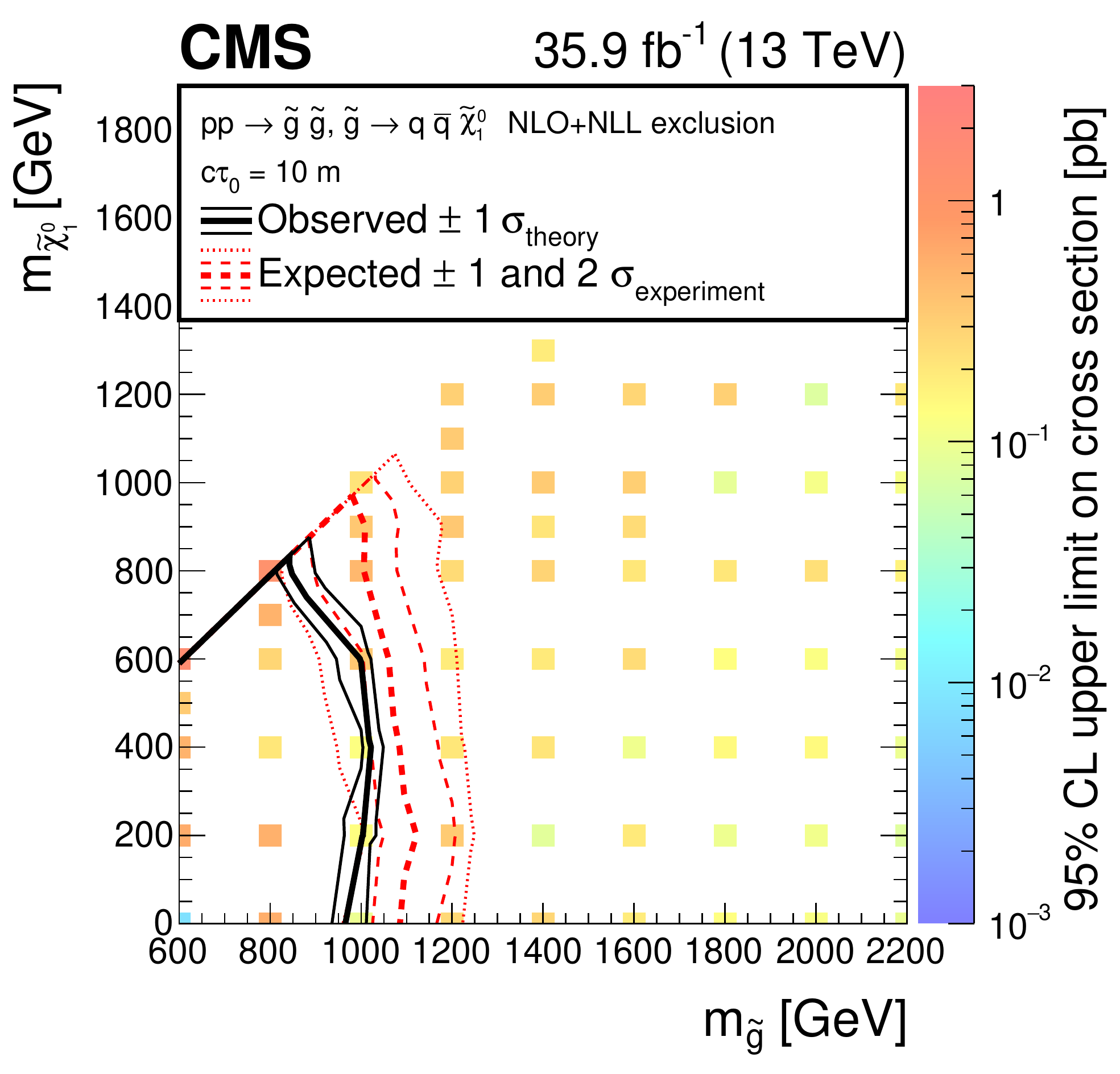}~
  \includegraphics[width=0.33\textwidth]{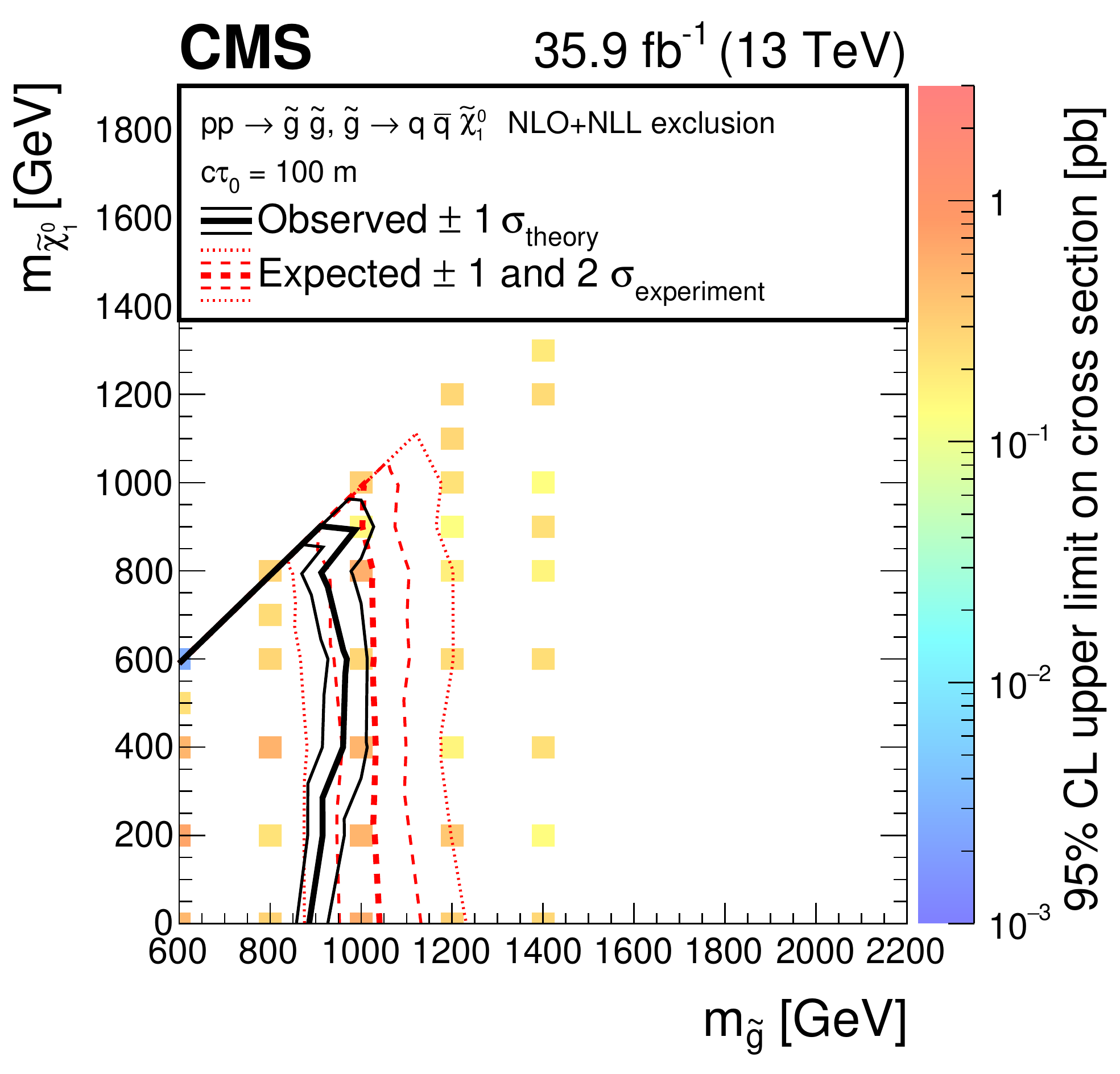}\\
  \caption{Observed upper limit in cross section at 95\% \CL (indicated
    by the colour scale) as a function of the $\PSg$ and \PSGczDo
    masses for simplified models that assume the production of pairs
    of long-lived gluinos that each decay via highly virtual
    light-flavour squarks to the neutralino and SM particles
    (T1qqqqLL). Each subfigure represents a different gluino lifetime:
    $\ctau = 1$ (upper left), 10 (upper centre), and $100\um$ (upper
    right); 1 (middle left), 10 (middle centre), and $100\unit{mm}$
    (middle right); and 1 (lower left), 10 (lower centre), and
    $100\unit{m}$ (lower right).  The thick (thin) black solid line
    indicates the observed excluded region assuming the nominal
    (${\pm}1$ standard deviation in theoretical uncertainty)
    production cross section. The red thick dashed (thin dashed and
    dotted) line indicates the median (${\pm}1$ and 2 standard
    deviations in experimental uncertainty) expected excluded region.
  }
  \label{fig:limits-ll}
\end{figure}

Figure~\ref{fig:limits-ll} summarizes the evolution of the search
sensitivity to the T1qqqqLL family of models as a function of
\ctau. Each subfigure presents the observed $\sigma_\text{UL}$ as a
function of $m_{\PSg}$ and $m_{\PSGczDo}$ for simplified models that
involve the production of gluino pairs. The excluded mass regions
based on the observed and expected values of $\sigma_\text{UL}$ are
also shown, along with contours determined under variations in
theoretical and experimental uncertainties. The top row of subfigures
cover the range $1 < \ctau < 100\um$ and demonstrate coverage
comparable to the T1qqqq prompt-decay scenario. A moderate improvement
in sensitivity for models with $1 \lesssim \ctau \lesssim 10\unit{mm}$
is observed because of the additional signal-to-background
discrimination provided by the \nb variable. The sensitivity is
reduced for models with lifetimes in the region $\ctau > 100\unit{mm}$
because of a lower acceptance for the jets from the gluino decay and
an increased reliance on jets from ISR. The coverage is independent of
\ctau beyond values of $10\unit{m}$ and comparable to the limiting
case of a metastable gluino.

A nonnegligible fraction of R-hadrons that traverse the muon chambers
before decaying are identified as muons by the PF algorithm. The
fraction is dependent on the R-hadron model and the choice of
parameters that affect the hadronization model and matter
interactions. The signal \ate is strongly dependent on \ctau due to
the muon veto employed by this search. Under these assumptions, the
excluded mass regions shown in Fig.~\ref{fig:limits-ll} weaken by
50--200\GeV for models with $\ctau \gtrsim 1\unit{m}$, with the
largest change occurring at $\ctau \approx 10\unit{m}$. The change is
negligible for models with \ctau below $1\unit{m}$.

Table~\ref{tab:limits} summarizes the strongest expected and observed
mass limits for each family of models. The simplified result based on
the 32-bin schema, summarized in Appendix~\ref{app:suppMat}, yields
limits on $\sigma_\text{UL}$ that are typically a factor ${\approx}2$
weaker than those obtained with the nominal result.

\begin{table}[!h]
  \topcaption{Summary of the mass limits obtained for each family of
    simplified models. The limits indicate the strongest observed
    mass exclusions for the parent SUSY particle (gluino or
    squark) and \PSGczDo.
  }
  \label{tab:limits}
  \centering
  \begin{tabular}{ lcc }
    \hline
    Model family                    & \multicolumn{2}{c}{Best mass limit [\GeVns{}]} \\ [0.5ex]
    \cline{2-3}
                                    & Gluino or squark & \PSGczDo                    \\ [0.5ex]
    \hline
    T2bb                            & 1050             & \ph{1}500                   \\
    T2tt                            & 1000             & \ph{1}400                   \\
    T2cc                            & \ph{1}500        & \ph{1}475                   \\
    T2qq\_8fold                     & 1325             & \ph{1}575                   \\
    T2qq\_1fold                     & \ph{1}675        & \ph{1}350                   \\
    T1bbbb                          & 1900             & 1150                        \\
    T1tttt                          & 1650             & \ph{1}850                   \\
    T1qqqq                          & 1650             & \ph{1}900                   \\
    T1qqqqLL (Metastable $\PSg$)    & \ph{1}900        & \NA                         \\
    T1qqqqLL ($\ctau = 1\unit{mm}$) & 1750             & 1000                        \\
    \hline
  \end{tabular}
\end{table}

\section{Summary}
\label{sec:summary}

A search for supersymmetry with the CMS experiment is reported, based
on a data sample of {\Pp\Pp} collisions collected in 2016 at $\sqrt{s} =
13\TeV$ that corresponds to an integrated luminosity of $35.9 \pm 0.9
\fbinv$. Final states with jets and significant missing transverse
momentum \ptvecmiss, as expected from the production and decay of
massive gluinos and squarks, are considered. Signal events are
categorized according to the number of reconstructed jets, the number
of jets identified as originating from bottom quarks, and the scalar
and vector sums of the transverse momenta of jets. The standard model
background is estimated from a binned likelihood fit to event yields
in the signal region and data control samples. The observed yields in
the signal region are found to be in agreement with the expected
contributions from standard model processes. Supplemental material is
provided to aid further interpretation of the result in
Appendix~\ref{app:suppMat}.

Limits are determined in the parameter spaces of simplified models
that assume the production and prompt decay of gluino or squark
pairs. The strongest exclusion bounds (95\% confidence level) for
squark masses are 1050, 1000, and 1325\GeV for bottom, top, and
mass-degenerate light-flavour squarks, respectively. The corresponding
mass bounds on the neutralino \PSGczDo from squark decays are 500,
400, and 575\GeV. The gluino mass is probed up to 1900, 1650, and
1650\GeV when the gluino decays via virtual states of the
aforementioned squarks. The strongest mass bound on the \PSGczDo from
the gluino decay is 1150\GeV.

Sensitivity to simplified models inspired by split supersymmetry is
also demonstrated. These models assume the production of long-lived
gluino pairs that decay to final states containing displaced jets and
\ptvecmiss from the undetected \PSGczDo particles. The long-lived
gluino, with an assumed proper decay length \ctau, is expected to
hadronize with SM particles and form a bound state known as an
R-hadron. The model-dependent matter interactions of R-hadrons are not
considered by default. The sensitivity of this search is only
moderately dependent on these matter interactions for models with
$\ctau \gtrsim 1\unit{m}$, while no dependence is found for models
with \ctau below 1\unit{m}. Models that assume a \PSGczDo mass of
100\GeV and gluino masses up to 1600\GeV are excluded for a proper
decay length \ctau below 0.1\unit{mm}. The bound on the gluino mass
strengthens to 1750\GeV at $\ctau = 1\unit{mm}$, before weakening to
900--1000\GeV for models with $\ctau > 10\unit{m}$.  For all values of
\ctau considered, the exclusion bounds on the gluino mass weaken to
about 1\TeV when the difference between the gluino and \PSGczDo mass
is small. The search provides coverage of the \ctau parameter space
for models involving long-lived gluinos, such as the region $\ctau
\lesssim 1\unit{mm}$, that is complementary to the coverage provided
by dedicated techniques at the LHC.

\begin{acknowledgments}
We congratulate our colleagues in the CERN accelerator departments for
the excellent performance of the LHC and thank the technical and
administrative staffs at CERN and at other CMS institutes for their
contributions to the success of the CMS effort. In addition, we
gratefully acknowledge the computing centres and personnel of the
Worldwide LHC Computing Grid for delivering so effectively the
computing infrastructure essential to our analyses. Finally, we
acknowledge the enduring support for the construction and operation of
the LHC and the CMS detector provided by the following funding
agencies: BMWFW and FWF (Austria); FNRS and FWO (Belgium); CNPq,
CAPES, FAPERJ, and FAPESP (Brazil); MES (Bulgaria); CERN; CAS, MoST,
and NSFC (China); COLCIENCIAS (Colombia); MSES and CSF (Croatia); RPF
(Cyprus); SENESCYT (Ecuador); MoER, ERC IUT, and ERDF (Estonia);
Academy of Finland, MEC, and HIP (Finland); CEA and CNRS/IN2P3
(France); BMBF, DFG, and HGF (Germany); GSRT (Greece); OTKA and NIH
(Hungary); DAE and DST (India); IPM (Iran); SFI (Ireland); INFN
(Italy); MSIP and NRF (Republic of Korea); LAS (Lithuania); MOE and UM
(Malaysia); BUAP, CINVESTAV, CONACYT, LNS, SEP, and UASLP-FAI
(Mexico); MBIE (New Zealand); PAEC (Pakistan); MSHE and NSC (Poland);
FCT (Portugal); JINR (Dubna); MON, RosAtom, RAS, RFBR and RAEP
(Russia); MESTD (Serbia); SEIDI, CPAN, PCTI and FEDER (Spain); Swiss
Funding Agencies (Switzerland); MST (Taipei); ThEPCenter, IPST, STAR,
and NSTDA (Thailand); TUBITAK and TAEK (Turkey); NASU and SFFR
(Ukraine); STFC (United Kingdom); DOE and NSF (USA).

\hyphenation{Rachada-pisek} Individuals have received support from the
Marie-Curie programme and the European Research Council and Horizon
2020 Grant, contract No. 675440 (European Union); the Leventis
Foundation; the A. P. Sloan Foundation; the Alexander von Humboldt
Foundation; the Belgian Federal Science Policy Office; the Fonds pour
la Formation \`a la Recherche dans l'Industrie et dans l'Agriculture
(FRIA-Belgium); the Agentschap voor Innovatie door Wetenschap en
Technologie (IWT-Belgium); the Ministry of Education, Youth and Sports
(MEYS) of the Czech Republic; the Council of Science and Industrial
Research, India; the HOMING PLUS programme of the Foundation for
Polish Science, cofinanced from European Union, Regional Development
Fund, the Mobility Plus programme of the Ministry of Science and
Higher Education, the National Science Center (Poland), contracts
Harmonia 2014/14/M/ST2/00428, Opus 2014/13/B/ST2/02543,
2014/15/B/ST2/03998, and 2015/19/B/ST2/02861, Sonata-bis
2012/07/E/ST2/01406; the National Priorities Research Program by Qatar
National Research Fund; the Programa Severo Ochoa del Principado de
Asturias; the Thalis and Aristeia programmes cofinanced by EU-ESF and
the Greek NSRF; the Rachadapisek Sompot Fund for Postdoctoral
Fellowship, Chulalongkorn University and the Chulalongkorn Academic
into Its 2nd Century Project Advancement Project (Thailand); the Welch
Foundation, contract C-1845; and the Weston Havens Foundation (USA).
\end{acknowledgments}

\bibliography{auto_generated}

\clearpage
\appendix
\section{Supplemental material\label{app:suppMat}}

\begingroup
\renewcommand*{\arraystretch}{1.2}
\begin{table}[!h]
  \topcaption{Summary of the nominal (\njet, \nb, \scalht, \mht)
    binning schema. Each entry (and the following entry, if present)
    signifies the lower (upper) bound of an \mht bin within a given
    (\njet, \nb, \scalht) bin. Unique or final entries represent \mht
    bins unbounded from above. A dash (\NA) signifies that the \scalht
    bin in a given (\njet, \nb) category is not used in the analysis,
    in which case counts in high-\scalht bins are integrated into the
    adjacent lower-\scalht bin. For monojet events, $\scalht \equiv
    \mht$. The \textit{a} denotes asymmetric \pt thresholds for the two
    highest \pt jets.
  }
  \label{tab:binning}
  \centering
  \resizebox{\textwidth}{!}{
    \begin{tabular}{rrlllll}
      \hline
      \njet      & \nb       & \multicolumn{5}{c}{\scalht [\GeVns{}]}                                                                       \\
      \cline{3-7}
                 &           & 200 & 400            & 600                       & 900                                  & 1200               \\
      \hline
      1          & 0         & 200 & 400 \ph{, 200} & 600 \ph{, 200} \ph{, 200} & 900 \ph{, 200} \ph{, 200} \ph{, 200} & \NA                \\
      1          & 1         & 200 & 400 \ph{, 200} & 600 \ph{, 200} \ph{, 200} & \NA                                  & \NA                \\
      ${\geq}2a$ & 0         & 200 & 200, 400       & 200, 400, 600             & 200, 900 \ph{, 200} \ph{, 200}       & \NA                \\
      ${\geq}2a$ & 1         & 200 & 200, 400       & 200, 400, 600             & 200, 900 \ph{, 200} \ph{, 200}       & \NA                \\
      ${\geq}2a$ & 2         & 200 & 200, 400       & 200, 400, 600             & 200, 900 \ph{, 200} \ph{, 200}       & \NA                \\
      ${\geq}2a$ & ${\geq}3$ & 200 & 200, 400       & 200, 400, 600             & \NA                                  & \NA                \\
      2          & 0         & 200 & 200, 400       & 200, 400, 600             & 200, 400, 600, 900                   & 200, 400, 600, 900 \\
      2          & 1         & 200 & 200, 400       & 200, 400, 600             & 200, 400, 600, 900                   & 200, 400, 600, 900 \\
      2          & 2         & 200 & 200, 400       & 200, 400, 600             & \NA                                  & \NA                \\
      3          & 0         & 200 & 200, 400       & 200, 400, 600             & 200, 400, 600, 900                   & 200, 400, 600, 900 \\
      3          & 1         & 200 & 200, 400       & 200, 400, 600             & 200, 400, 600, 900                   & 200, 400, 600, 900 \\
      3          & 2         & 200 & 200, 400       & 200, 400, 600             & 200, 400, 600, 900                   & 200, 400, 600, 900 \\
      3          & 3         & 200 & 200, 400       & 200, 400, 600             & \NA                                  & \NA                \\
      4          & 0         & \NA & 200, 400       & 200, 400, 600             & 200, 400, 600, 900                   & 200, 400, 600, 900 \\
      4          & 1         & \NA & 200, 400       & 200, 400, 600             & 200, 400, 600, 900                   & 200, 400, 600, 900 \\
      4          & 2         & \NA & 200, 400       & 200, 400, 600             & 200, 400, 600, 900                   & 200, 400, 600, 900 \\
      4          & ${\geq}3$ & \NA & 200, 400       & 200, 400, 600             & 200, 400, 600, 900                   & \NA                \\
      5          & 0         & \NA & 200, 400       & 200, 400, 600             & 200, 400, 600 \ph{, 200}             & 200, 400, 600, 900 \\
      5          & 1         & \NA & 200, 400       & 200, 400, 600             & 200, 400, 600 \ph{, 200}             & 200, 400, 600, 900 \\
      5          & 2         & \NA & 200, 400       & 200, 400, 600             & 200, 400, 600 \ph{, 200}             & 200, 400, 600, 900 \\
      5          & 3         & \NA & 200, 400       & 200, 400, 600             & 200, 400, 600 \ph{, 200}             & \NA                \\
      5          & ${\geq}4$ & \NA & 200, 400       & \NA                       & \NA                                  & \NA                \\
      ${\geq}6$  & 0         & \NA & 200 \ph{, 200} & 200, 400 \ph{, 200}       & 200, 400, 600 \ph{, 200}             & 200, 400, 600, 900 \\
      ${\geq}6$  & 1         & \NA & 200 \ph{, 200} & 200, 400 \ph{, 200}       & 200, 400, 600 \ph{, 200}             & 200, 400, 600, 900 \\
      ${\geq}6$  & 2         & \NA & 200 \ph{, 200} & 200, 400 \ph{, 200}       & 200, 400, 600 \ph{, 200}             & 200, 400, 600, 900 \\
      ${\geq}6$  & 3         & \NA & 200 \ph{, 200} & 200, 400 \ph{, 200}       & 200, 400, 600 \ph{, 200}             & 200, 400, 600, 900 \\
      ${\geq}6$  & ${\geq}4$ & \NA & 200 \ph{, 200} & \NA                       & \NA                                  & \NA                \\
      \hline
    \end{tabular}
  }
\end{table}
\endgroup

\begingroup
\renewcommand*{\arraystretch}{1.1}
\begin{table}[!t]
  \topcaption{Observed counts of candidate signal events and SM
    expectations determined from the CR-only fit using the simplified
    binning schema, as a function of \njet, \nb, and \mht. All counts
    are integrated over \scalht. The uncertainties include both
    statistical and systematic contributions.
    The \textit{a} denotes asymmetric \pt thresholds for the two highest \pt
    jets.
  }
  \label{tab:simplified}
  \centering
  \begin{tabular}{rrlr@{}lr@{}lr@{}lr@{}l}
    \hline
    \njet          & \nb       &      & \multicolumn{8}{c}{\mht [\GeVns{}]}                                                                        \\
    \cline{4-11}
                   &           &      & 200        &                   & 400       &                & 600   &               & 900   &              \\
    \hline
    =1, ${\geq}2a$ & 0         & Data & 411\,184   &                   & 11\,448   &                & 1116  &               & 111                  \\
                   &           & SM   & $360\,000$ & $\,\pm\, 38\,000$ & $10\,000$ & $\,\pm\, 1400$ & $910$ & $\,\pm\, 170$ & $107$ & $\,\pm\, 28$ \\[0.2ex]
    =1, ${\geq}2a$ & ${\geq}1$ & Data & 31\,174    &                   & 769       &                & 105   &               & 7                    \\
                   &           & SM   & $25\,500$  & $\,\pm\, 2500$    & $649$     & $\,\pm\, 91$   & $69$  & $\,\pm\, 13$  & $6$   & $.4 \pm 1.8$ \\[0.2ex]
    =2, =3         & =0, =1    & Data & 66\,955    &                   & 5946      &                & 903   &               & 100                  \\
                   &           & SM   & $58\,000$  & $\,\pm\, 11\,000$ & $5400$    & $\,\pm\, 1100$ & $860$ & $\,\pm\, 220$ & $113$ & $\,\pm\, 41$ \\[0.2ex]
    =2, =3         & ${\geq}2$ & Data & 1045       &                   & 70        &                & 6     &               & 0                    \\
                   &           & SM   & $870$      & $\,\pm\, 130$     & $56$      & $.9 \pm 9.4$   & $7$   & $.1 \pm 1.7$  & $1$   & $.0 \pm 0.4$ \\[0.2ex]
    =4, =5         & =0, =1    & Data & 9546       &                   & 1734      &                & 315   &               & 44                   \\
                   &           & SM   & $10\,500$  & $\,\pm\, 1100$    & $1880$    & $\,\pm\, 310$  & $319$ & $\,\pm\, 71$  & $40$  & $\,\pm\, 14$ \\[0.2ex]
    =4, =5         & ${\geq}2$ & Data & 1012       &                   & 93        &                & 4     &               & 3                    \\
                   &           & SM   & $970$      & $\,\pm\, 110$     & $81$      & $\, \pm 11$    & $8$   & $.4 \pm 1.7$  & $1$   & $.2 \pm 0.4$ \\[0.2ex]
    ${\geq}6$      & =0, =1    & Data & 758        &                   & 141       &                & 33    &               & 5                    \\
                   &           & SM   & $910$      & $\,\pm\, 180$     & $167$     & $\,\pm\, 76$   & $33$  & $\,\pm\, 25$  & $4$   & $.2 \pm 5.0$ \\[0.2ex]
    ${\geq}6$      & ${\geq}2$ & Data & 197        &                   & 14        &                & 3     &               & 0                    \\
                   &           & SM   & $189$      & $\,\pm\, 40$      & $16$      & $.9 \pm 4.9$   & $2$   & $.1 \pm 1.2$  & $0$   & $.2 \pm 0.2$ \\
    \hline
  \end{tabular}
\end{table}
\endgroup

\begin{figure}[!b]
  \centering
  \includegraphics[width=0.6\textwidth]{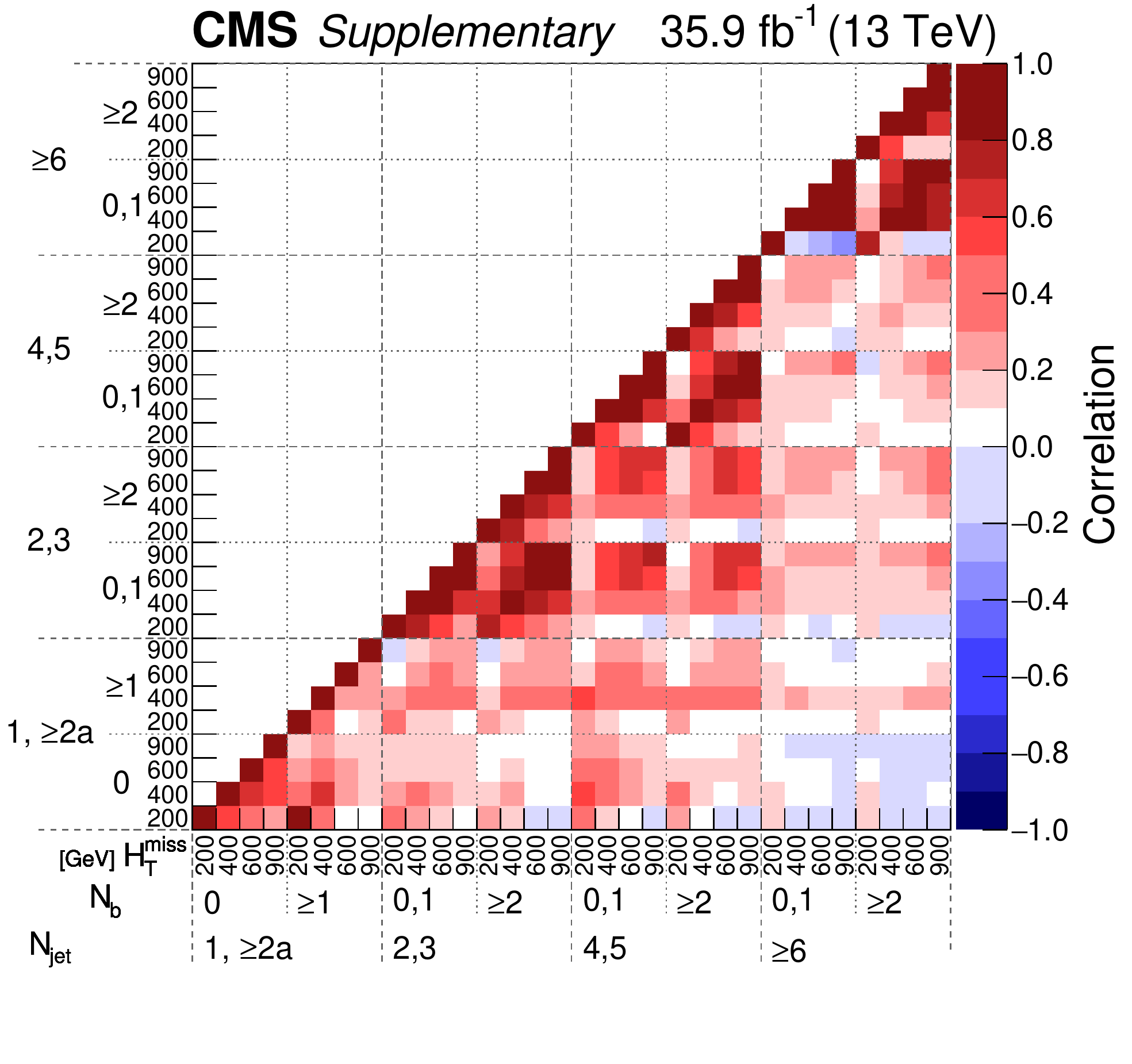}
  \caption{Correlation matrix for the SM background estimates
    determined from the CR-only fit using the simplified binning
    schema defined in Table~\ref{tab:simplified}.}
  \label{fig:correlation}
\end{figure}
\cleardoublepage \section{The CMS Collaboration \label{app:collab}}\begin{sloppypar}\hyphenpenalty=5000\widowpenalty=500\clubpenalty=5000\input{SUS-16-038-authorlist.tex}\end{sloppypar}
\end{document}

%% file: SUS-16-038-authorlist.tex
\textbf{Yerevan Physics Institute,  Yerevan,  Armenia}\\*[0pt]
A.M.~Sirunyan, A.~Tumasyan
\vskip\cmsinstskip
\textbf{Institut f\"{u}r Hochenergiephysik,  Wien,  Austria}\\*[0pt]
W.~Adam, F.~Ambrogi, E.~Asilar, T.~Bergauer, J.~Brandstetter, E.~Brondolin, M.~Dragicevic, J.~Er\"{o}, A.~Escalante Del Valle, M.~Flechl, M.~Friedl, R.~Fr\"{u}hwirth\cmsAuthorMark{1}, V.M.~Ghete, J.~Grossmann, J.~Hrubec, M.~Jeitler\cmsAuthorMark{1}, A.~K\"{o}nig, N.~Krammer, I.~Kr\"{a}tschmer, D.~Liko, T.~Madlener, I.~Mikulec, E.~Pree, N.~Rad, H.~Rohringer, J.~Schieck\cmsAuthorMark{1}, R.~Sch\"{o}fbeck, M.~Spanring, D.~Spitzbart, A.~Taurok, W.~Waltenberger, J.~Wittmann, C.-E.~Wulz\cmsAuthorMark{1}, M.~Zarucki
% ...
\vskip\cmsinstskip
% ...
1:~~Also at Vienna University of Technology, Vienna, Austria\\
% ...
